\documentclass[prd,twocolumn,nofootinbib,aps,tightenlines,preprintnumbers,notitlepage,longbibliography,superscriptaddress,floats,reprint,showpacs]{revtex4-1}
\usepackage{graphicx}
\usepackage{amsmath}
\usepackage{amsfonts}
\usepackage[colorlinks=true,citecolor=blue,urlcolor=cyan,linkcolor=black]{hyperref}
\usepackage{scalerel}

\newcommand{\nhat}{\hat{ \mathbf{n}}}

\usepackage{amsmath}
\usepackage{amsfonts}

\usepackage{tabularx}
\newcolumntype{C}{>{\centering\arraybackslash}X}
\newcolumntype{R}{>{\raggedleft\arraybackslash}X}

\usepackage[usenames, dvipsnames]{color}
\usepackage[normalem]{ulem}
\usepackage{xcolor}

\newcommand{\aap}{Astronomy and Astrophysics}
\newcommand{\jcap}{Journal of Cosmology and Astroparticle Physics}

\newcommand{\dd}{{\rm d}}
\newcommand{\bl}{{\boldsymbol{\ell}}}

\def\nhat{\hat{\mathbf{n}}}

\newcommand{\bL}{\boldsymbol{L}}

\newcommand{\bx}{\boldsymbol{x}}

\newcommand{\be}{\begin{eqnarray}}
\newcommand{\non}{\nonumber \\}
\newcommand{\ee}{\end{eqnarray}}

\definecolor{colorA}{HTML}{1E90FF}
\definecolor{colorB}{HTML}{228B22}
\definecolor{colorC}{HTML}{FF7F00}
\definecolor{colorD}{HTML}{4B0082}
\definecolor{colorE}{HTML}{B22222}

\definecolor{lgreen}{HTML}{32CD32}
\definecolor{lgray}{HTML}{D3D3D3}
\definecolor{dblue}{HTML}{1E90FF}
\definecolor{dblue}{HTML}{1E90FF}
\definecolor{orange}{HTML}{FF4500}
\definecolor{indigo}{HTML}{4B0082}

\definecolor{teal}{HTML}{008080}
\definecolor{firebrick}{HTML}{B22222}
\definecolor{salmon}{HTML}{FA8072}
\definecolor{darkgreen}{HTML}{006400}

\newcommand{\illb}{Department of Physics, University of Illinois Urbana-Champaign, 1110 W. Green Street, Urbana, IL 61801, USA}

\newcommand{\perimeter}{Perimeter Institute for Theoretical Physics, 31 Caroline St N, Waterloo, ON N2L 2Y5, Canada}

\def\aaps{\ref@jnl{A\&AS}}
\providecommand\mnras{Monthly Notices of the Royal Astronomical Society}

\graphicspath{ {plots/} }

\begin{document}

\title{Probing the optical depth with galaxy number counts}

\author{Selim~C.~Hotinli}
\affiliation{\perimeter}

\author{Gilbert~P.~Holder}
\affiliation{\illb}

\begin{abstract}

We study the prospects for measuring the cosmological distribution and abundance of ionized electrons in the intergalactic medium using galaxy surveys. Optical light from distant galaxies is subject to Thomson screening by intervening electrons which distorts the observed galaxy number density, similar to the effect of weak gravitational lensing magnification. We construct an estimator for the optical-depth fluctuations from the statistical anisotropies of galaxy number counts induced by the spatially-varying optical-depth field. We find near-future galaxy surveys can detect this signal at signal-to-noise above $\sim10$ depending on galaxy survey specifications. We highlight various science cases for the measurement of optical-depth fluctuations. 

\end{abstract}

\maketitle

\section{Introduction}\label{sec:Intro}\vspace*{-0.2cm}

The optical light arriving from far-away galaxies interacts with intervening large-scale structure (LSS) through scattering and gravitational lensing effects. Such interactions alter the number counts of galaxies observed with flux-limited surveys, obscuring the measurement of underlying galaxy-clustering signal, as well as providing an additional window into probing distributions of matter and baryons throughout the cosmic history. As an array of galaxy surveys including DESI~\citep{DESI:2016fyo}, LSST~\citep{LSSTDarkEnergyScience:2018jkl}, Euclid~\citep{2011arXiv1110.3193L} and Roman~\citep{2015arXiv150303757S} stand to provide the most precise galaxy data sets up to date, these effects will be measured to high significance for the first time in the next years. Here we study the prospects of detecting Thomson screening by free electrons in the intergalactic medium (IGM) from galaxy number counts.

Thomson screening modifies the incoming flux by a factor $\exp(-\tau)$ where $\tau$ is the optical depth. This leads to the observed flux being either enhanced or reduced in a  galaxy survey, depending on the spatial fluctuations in $\tau$, similar to reddening due to dust~\citep[e.g.][]{2010MNRAS.405.1025M, 2015ApJ...813....7P} and lensing magnification effects~\citep[e.g.][]{2002ApJ...580L...3J, 2005ApJ...633..589S, Hoekstra:2008db, 2009A&A...507..683H, 2013MNRAS.429.3230H, 2011PhRvD..84j3004V, 2012ApJ...744L..22S, 2019MNRAS.482..785S, 2021MNRAS.504.1452V}. At linear order in cosmological perturbations, these effects add modulations to the observed number counts of galaxies in a flux-limited survey that will be proportional to $\tau$. At next-to-leading order, these modulations drive the galaxy number count power-spectra away from being statistically stationary: there will be spatially varying power depending on the large-scale fluctuations in $\tau$. {This effect is similar to lensing, where at linear order, galaxy number count fluctuations see modulations proportional to lensing convergence $\kappa$, while at next-to-leading order, galaxy power spectra similarly become statistically non-stationary and anisotropic due to lensing.}

The prospects of using these distortions to measure or reconstruct large-scale cosmological fluctuations has been studied before in the literature for lensing reconstruction~\citep[e.g.][]{Nistane:2022xuz,Wenzl:2023wsd,Ma:2023kyc}, where forecasts suggest a high-significance detection could be achievable with current and near-future data sets. Reconstructing large-scale optical-depth fluctuations due to electrons (an effect roughly $\sim20$ times smaller) using similar techniques has not yet been studied. We will show that upcoming galaxy surveys such as LSST will have the statistical power to detect this signal at signal-to-noise level reaching beyond $\sim10$, and highlight possible science drivers for these measurements.

Fluctuations of optical depth carry valuable cosmological and astrophysical information. On large scales, cross-correlations between matter and baryon (or electron) fluctuations probe initial conditions, interactions and dynamics during early Universe, such as primordial isocurvature~\citep[e.g.][]{2009PhRvD..80f3535G,2011PhRvD..84l3003G,2011PhRvL.107z1301G,2014PhRvD..89b3006G,2016PhRvL.116t1302S,2017PhRvD..96h3508S,2019MNRAS.485.1248S,2019PhRvD.100f3503H,2019PhRvD.100j3528H,2020JCAP...07..049B,2021PhRvD.104f3536H,2023PhRvD.107d3504K,2023JCAP...08..051B,2023arXiv231118121V}, a smoking-gun signature of inflationary models that are different than the single-field slow-roll inflation~\citep{Linde:1996gt,Lyth:2002my,Langlois:2000ar,Sasaki:2006kq}, as well as other deviations from the standard cosmological paradigm, including primordial magnetic fields~\citep[e.g.][]{Subramanian:2015lua,Carrilho:2019qlb,Flitter:2023xql} and primordial black holes~\citep[e.g.][]{Passaglia:2021jla}, for example. On scales corresponding to the circumgalactic medium and the intergalactic medium (IGM), precise measurements of free electron abundance might give hints about the strength of baryonic feedback which can push gas away from the halo centers. Such hints could improve our current understanding of galaxy formation~\citep[e.g.][]{Faucher-Giguere:2023lgr,2021MNRAS.504.5131L}, further elucidate the `missing baryon' problem~\citep[e.g.][]{deGraaff:2017byg}, or point to the influence of baryons on matter perturbations beyond halo scales, with implications for the so-called `$S_8$ tension'~\citep[e.g.][]{Amon:2022azi}. Small-scale baryon isocurvature fluctuations can also be a probe of initial conditions~\citep[e.g.][]{Lee:2021bmn}.

Furthermore, optical depth is a tracer of the ionization history. In particular, at redshifts $2<z<4$ accessible to upcoming galaxy surveys, the mean ionization fraction probes the ionization of the second electron in helium (hereafter `helium reionization')~\citep{Wyithe:2002qu,Faucher-Giguere:2008ksu, Madau:2015cga,McQuinn:2012bq,Worseck:2014gva,Furlanetto:2007mg,Worseck:2011qk,Sokasian:2001xh,Compostella:2013zya,Oh:2000sg,Furlanetto:2007gn,Furlanetto:2008qy,Dixon:2009xa,LaPlante:2016bzu,Caleb:2019apf,Linder:2020aru,Meiksin:2011bq,Compostella:2014joa,Eide:2020xyi,UptonSanderbeck:2020zla,Bhattacharya:2020rtf,Villasenor:2021ksg,Meiksin:2010rv,Gotberg:2019uhh,LaPlante:2015rea,Syphers:2011uw,Dixon:2013gea,LaPlante:2017xzz,Lau:2020chu,Hotinli:2022jna,Hotinli:2022jnt,Caliskan:2023yov}. Characterising helium reionization has great significance for understanding galaxy formation, quasar activity and cosmology~\cite[e.g.][]{Masters2012,2013ApJ...773...14R,2013ApJ...768..105M,McGreer:2017myu,2022ApJ...928..172P}. Since photons emitted by the first stars (sourcing reionization of hydrogen) are not energetic enough to fully ionize intergalactic helium, helium reionization occurs only after the emergence of a substantial number of quasars or active galactic nuclei. As a result, the history of helium reionization strongly depends on the properties of quasars, such as their {luminosity function}, {accretion mechanisms} and other astrophysics, clustering, variability, lifetimes, as well as the general growth and evolution of {super-massive black holes}~\citep[e.g.][]{Shen:2014rka,Hopkins:2006vv,2017ApJ...847...81S,Inayoshi:2019fun}. Since essentially all of the helium in the Universe is ultimately doubly ionized, the total change in the ionization fraction is also a measure of the primordial helium abundance $Y_p$, a sensitive probe of {big bang nucleosynthesis}. 

Finally, the distortions to observed galaxy number density due to optical depth can also bias the measurement of lensing. As the ability of near-future high-significance measurements of galaxy--galaxy-lensing cross-correlation to probe cosmological parameters will depend on the mitigation or modelling of lensing effects on the galaxy counts~\citep[see e.g.][and references therein]{Wenzl:2023wsd}, accounting for optical depth could play an important role in fully exploiting the statistical power of the upcoming galaxy survey data. 

This paper is organised as follows. We show the effect of Thomson screening and weak lensing on the galaxy number density measurements from flux-limited galaxy surveys in Sec.~\ref{sec:Num_count_distortions} (with supporting calculations presented in Appendix~\ref{sec:galaxy_fourier}). We define estimators for lensing and optical-depth fluctuations from variation of small-scale galaxy number density in Sec.~\ref{sec:estimators}. These estimators are derived explicitly in Appendix~\ref{app:estimators}. We highlight biases on these estimators in Appendix~\ref{sec:estimators_biases}. We present our results on the detection prospects of the optical depth in Sec.~\ref{sec:results}. In the same section, we also discuss the prospects of probing mean free electron abundance as a function of redshift. We conclude in Sec.~\ref{sec:discussion} with a discussion, and highlight upcoming work. We study the sensitivity of our results on some of our analysis choices in  Appendices~\ref{app:mag_nias}~and~\ref{sec:appendix}. In what follows we will assume for our forecasts that the galaxy distribution can be treated as Gaussian.

\vspace*{-0.2cm}
\section{Number count distortions}\label{sec:Num_count_distortions}\vspace*{-0.3cm}

\subsection{Modified observed flux}\label{sec:Anisotropic_flux}\vspace*{-0.3cm}

The observed flux can be written as
\be
F(\bx) = F_0(\bx)\, {\mu(\bx)}\, {e^{-\tau(\bx)}}\,,
\ee
where we use $\bx\equiv\nhat,z$ to indicate the coordinate in redshift space. Here, $z$ is redshift, $\nhat$ is the line-of-sight direction, $\mu(\bx)$ parametrises the effect of weak gravitational lensing on the measured flux, $\tau(\bx)$ is the optical depth, and $F_0(\bx)$ is the observed flux in the absence of Thomson screening and lensing effects.

The physical number of sources per redshift per solid angle at a given redshift and flux limit $F_*$ satisfy 
\be
\begin{split}
N (>F_*|\,z)&= \frac{1}{{\mu(\bx)}}N_{0}(>F_0|\,z)\\
&= \frac{1}{{\mu(\bx)}}N_{0}\Big(>\frac{F_*}{{\mu(\bx)}{e^{-\tau(\bx)}}}\Big|\,z\Big)\,,
\end{split}
\ee
where we define
\be
N_{0}(>F|z)=A\, F^{-\alpha(z)} n(z)\,,
\ee
with $A\,F^{-\alpha(z)}$ corresponding to a power-law model for the flux--number-count relation, and $n(z)$ is the redshift probability distribution of galaxies. Here $\alpha(z)$ is the change in the galaxy number density with respect to the luminosity cut at fixed redshift, defined as 
\be
\alpha(z)=-\frac{\partial\ln N(F_\star|\,z)}{\partial \ln F_\star}=-\frac{\partial n_g(L_\star|\,z)}{\partial \ln L_\star}\,,
\ee
where $n_g(L|\,z)$ is the comoving number density at the source, and $L$ is the luminosity cut satisfying $L_\star(z)=4\pi D_L(z)^2F_\star$, where $D_L(z)$ denotes the luminosity distance and $F_\star$ is the corresponding flux cut. The parameter $\alpha(z)$ is related to the magnification bias with a constant factor as $2\alpha(z)=5s(z)$ where $s(z)$ is the parameter used to study magnification bias when expressed in terms of magnitudes~\citep{2011PhRvD..84j3004V,2012ApJ...744L..22S}. In what follows we use $\alpha(z)$.

By modelling the observed flux with the Thomson screening and lensing effects, we find
\be
N (>F|\,z)={\mu(\bx)^{\alpha(z)-1}} {e^{-\alpha(z)\tau(\bx)}} A\,F_0^{-\alpha(z)}\,n(z)\,,
\ee
such that the spatial fluctuations in the number counts satisfy 
\begin{equation}
\begin{split}
\frac{N (>F|\,z)}{N_{0} (>F|\,z)}&={\mu(\bx)^{\alpha(z)-1}} {e^{-\alpha(z)\tau(\bx)}} \\
& \simeq\!1\!+\!{2[\alpha(z)\!-\!1]\kappa(\bx)}\!-\!{\alpha(z)\tau(\bx)}\,,
\end{split}
\end{equation}
where on the second line we have taken the weak lensing limit, $\mu(\bx)\simeq1 + 2\,\kappa(\bx)$, where $\kappa(\bx)$ is the lensing convergence. 

\vspace*{-0.2cm}
\subsection{Galaxy number counts}\label{sec:galaxy_no_count}\vspace*{-0.3cm}

\begin{figure*}[t!]
    \centering
    \includegraphics[width=\textwidth]{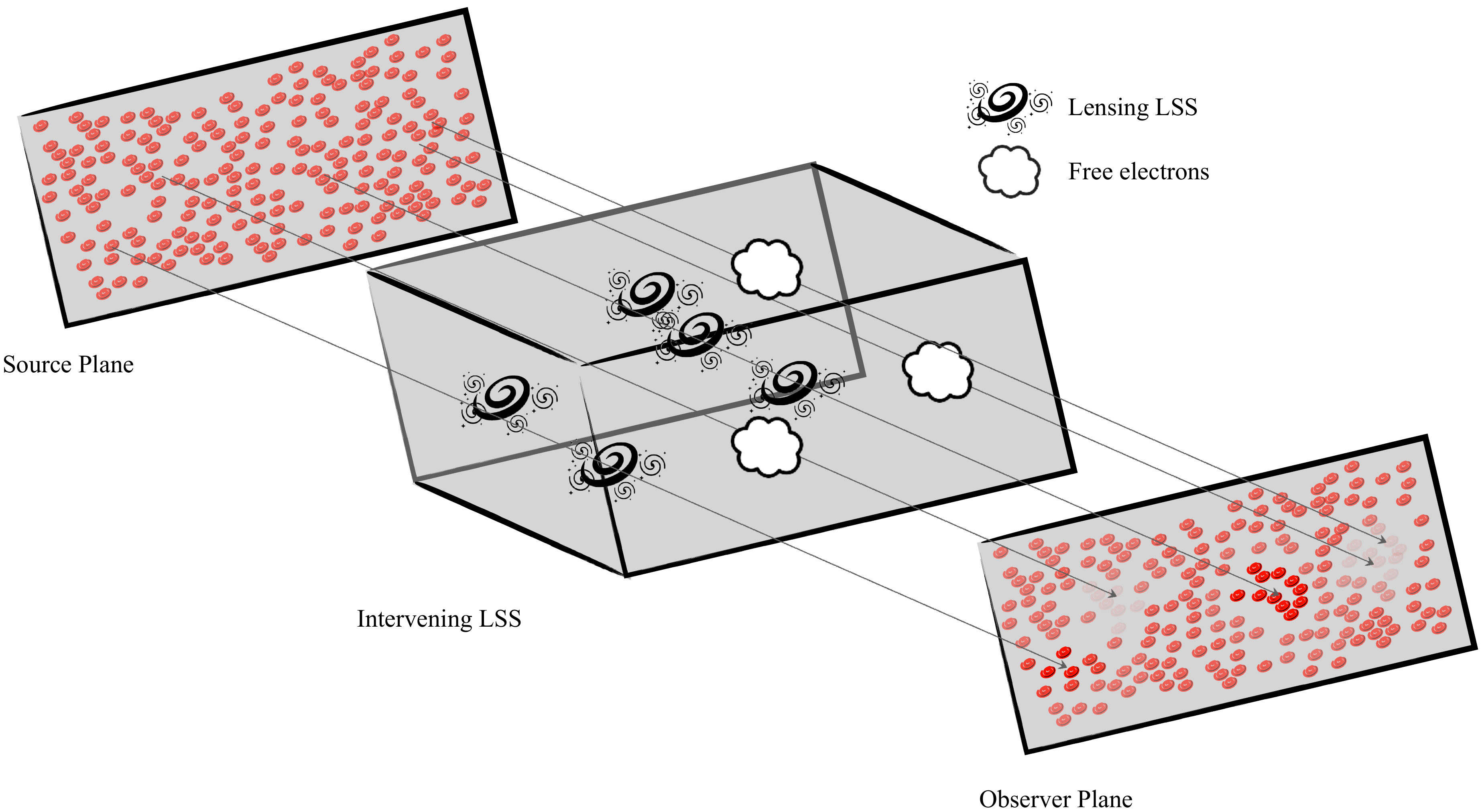}
    \vspace*{-0.5cm}
    \caption{\textit{Cartoon demonstrating the effect of magnification bias on galaxy survey observations.} Consider a source plane consisting of {identical} galaxies. Thomson screening due to free electrons in the IGM reduce the observed flux from some of these galaxies, making them more difficult to observe. At the same time, lensing magnification due to large-scale structure can enhance the flux from these galaxies, making them brighter and easier to observe. As a result, the luminosity of galaxies on the observer plane acquire spatially-varying modifications, tracing the distribution of intervening large-scale structure. In addition to these effects, note that lensing also induces shear-like distortions to observed galaxy number counts, which we are not shown in this plot.}
    \vspace*{-0.4cm}
    \label{fig:magnification_cartoon}
\end{figure*}

Building up on the previous section, the fluctuations in the galaxy number counts at leading order in perturbation theory (taking into account lensing and Thomson screening) can be written as
\be\label{eq:galaxy_realspace}
\begin{split}
\Delta_g(\bx)\!=\!\tilde{\Delta}_g(\bx)
+\delta\Delta^{\rm Lens}_g(\bx)+\delta\Delta^{\rm Thomson}_g(\bx)\,,
\end{split}
\ee
where
\be
\tilde{\Delta}_g(\bx)=b_g\delta(\bx)-\mathcal{H}^{-1}\nhat\cdot\boldsymbol{\nabla}(\nhat\cdot\boldsymbol{v})\,,
\ee 
is the sum of biased density fluctuations and the redshift-space distortions (RSDs). Here, $b_g$ is the galaxy bias, $\mathcal{H}$ is the Hubble rate and $\boldsymbol{v}$ is the bulk velocity. 

At leading order in galaxy density, the lensing contribution can be written as a combination of magnification and density-deflection effects, 
\begin{equation}\label{eq:lensing_realspace}
\begin{split}
&\delta\Delta^{\rm Lens}_g(\bx)\\
&\!=
\underbrace{\!-2[1\!-\!\alpha(z)]\kappa(\bx)[1 \!+\!\tilde{\Delta}_g(\bx)]}_{\rm magnification}\!+\underbrace{\!\nabla^a\phi(\bx)\nabla_a\tilde{\Delta}_g(\bx)}_{\rm density~deflection}
\end{split}
\end{equation}
where $2\,\kappa(\bx) = -\nabla^2\phi(\bx)$ and $\phi(\bx)$ is the projected lensing potential and gradients are defined in the transverse direction. The contribution from Thomson screening is analogous to lensing magnification, satisfying
\begin{equation}\label{eq:tau_realspace}
\delta\Delta^{\rm Thomson}_g(\bx)=\!-\alpha(z)\,\tau(\bx)[1\!+\!\tilde{\Delta}_g(\bx)]\,.
\end{equation}

We sketch the effects of Thomson screening and lensing magnification on galaxy observations in Fig.~\ref{fig:magnification_cartoon}. Here the source plane contains a distribution of identical galaxies. The intervening large-scale structure magnifies some of these galaxies, making them appear brighter, while Thomson screening leads to the opposite effect. This leads to spatially-varying modifications in the observed galaxy fluxes, which trace the underyling large-scale structure. For a flux-limited galaxy survey, these effects induce spatial variations in the galaxy number counts, as previously non-observable galaxies become brighter and observable due to lensing, and some observable galaxies become dimmer and non-observable due to Thomson screening, depending on the cosmological distribution of lensing mass and free electrons.

In addition to these effects, lensing also induces gravitational lensing deflections that modify both the shapes and the observed positions of galaxies. The shape distortions are measured as cosmic shear, while the deflected positions lead to spatial variations in the statistics of the density field \cite{Nistane:2022xuz}.

\begin{figure}[t!]
    \centering
    \includegraphics[width=0.49\textwidth]{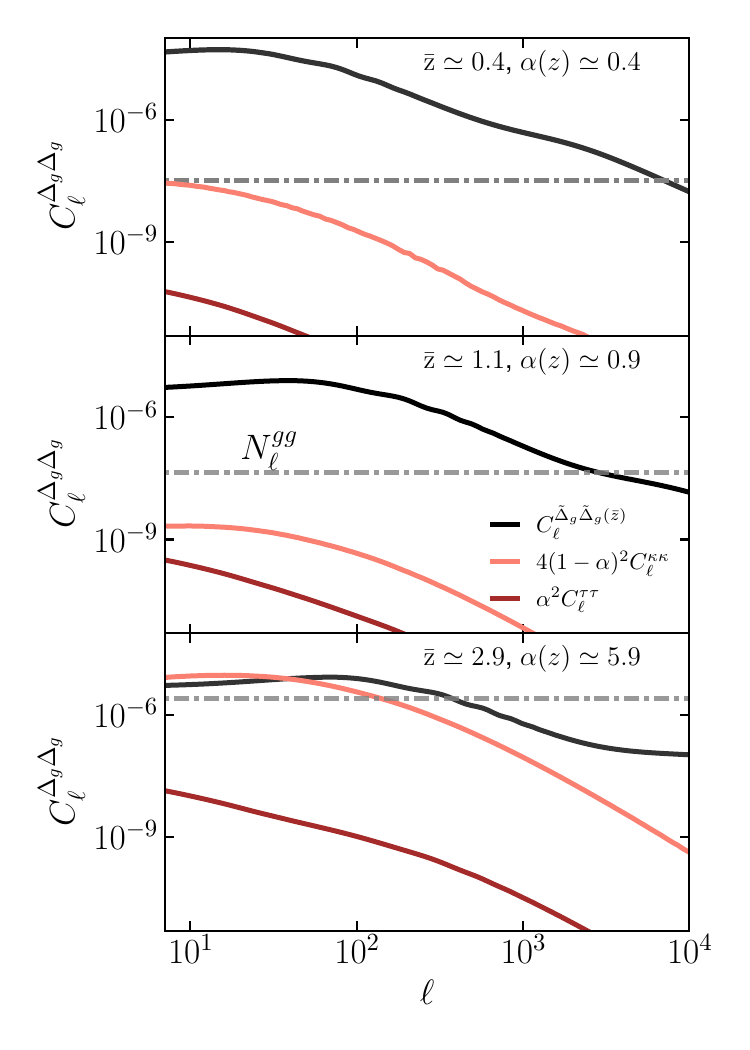}
    \vspace*{-0.7cm}
    \caption{\textit{{Contributions to the galaxy number count power spectra from lensing and Thomson screening.}} The top, middle and bottom panels correspond to redshift bins centered at redshifts $\bar{z}\simeq\{0.4,1.1,2.9\}$, respectively, with equal width in comoving distance (approximately $400\,$Mpc). The black solid lines correspond to galaxy number count power spectra in the absence of lensing and Thomson screening. The brown (light-red) lines correspond to the contribution from lensing (Thomson screening) auto-correlation. The dot-dashed gray lines correspond to shot noise anticipated from LSST survey `gold sample', satisfying $m_{\rm lim}=25$. We find at lower redshifts lensing dominates the corrections to the galaxy signal. The contributions from lensing and Thomson scattering becomes comparable at $\bar{z}\simeq1.1$. At higher redshifts, lensing again dominates, becoming comparable to the underlying galaxy signal.}  \label{fig:clgg_biases}
    \vspace*{-0.5cm}
\end{figure}

In Fig.~\ref{fig:clgg_biases} we show the effects of lensing and Thomson screening on galaxy number count power-spectra for a $m_{\rm lim}=25$ LSST-like survey (matching the LSST `gold sample' catalog). The galaxy power spectra are defined as 
\be\label{eq:galaxy_powerspec}
\langle\Delta_g(\bl,z)\Delta_g(\bl',z')\rangle=(2\pi)^2\delta_D^2(\bl\!+\!\bl') C_\ell^{\Delta_g\Delta_g(z,z')}\,,
\ee 
where $\bl$ is the Fourier wavevector, $\ell\equiv|\bl|$, $\delta_D^2$ is the 2-dimensional Dirac-delta function, and $\Delta_g(\bl,z)$ is the Fourier transform of galaxy number count fluctuations defined in Eq.~\eqref{eq:galaxy_realspace}. We calculate this signal explicitly in Appendix~\ref{sec:galaxy_fourier}. The three panels correspond to three of the 16 redshift bins we consider in this paper within $z\in[0.2,5.0]$, with equal sizes in comoving distance (approximately $400\,$Mpc); the three bins shown are centered at $\bar{z}\simeq\{0.4,\,1.1,\,2.9\}$ with the magnification bias parameters $\alpha(\bar{z})\simeq\{0.4,\,0.9,\,5.9\}$.

Overall, we find that lensing dominates the additional contributions to the  galaxy number count power spectrum as compared to Thomson screening, with the exception of the middle panel where $\alpha\simeq0.9$ and $\bar{z}\sim1.1$. At this redshift, both lensing and Thomson screening have a comparable effect on the galaxy number counts, since the lensing contribution depends on the magnification bias, in this case $1-\alpha(\bar{z})\sim0.1$; while the effect from Thomson screening is proportional to $\alpha(\bar{z})$. This shows the possible importance of accounting for Thomson screening in upcoming galaxy catalogues for values of the magnification bias parameter $\alpha(z)$ near unity. At redshifts below $z\sim1.1$, we find the total correction to the galaxy power spectra is over two orders of magnitude smaller compared to the underlying galaxy signal, while at higher redshifts $z\sim3$ and large scales, we find the lensing effect can be greater. Note that these results depend strongly on the value of $\alpha(z)$, which we demonstrate in Appendix~\ref{app:mag_bias_gal}.
In general, the effects on the power spectra are small, but we will show in the next section that there is more than simply the power spectra. This is similar to the case of CMB lensing, where lensing can be seen to slightly smooth the acoustic peaks but there is also a large amount of information that can be extracted from the way that lensing changes the local statistics of the CMB field.

\vspace*{-0.2cm}
\section{Estimators}\label{sec:estimators}\vspace*{-0.3cm}

In this section we introduce estimators that aim to reconstruct the underlying optical-depth and lensing fluctuations from their imprint on the galaxy number count distributions. Since galaxy number counts are modulated both at linear and quadratic orders in density, we can write estimators using each of these effects, and calculate the minimum-variance reconstruction noise from these, by adding them in inverse quadrature. 

The linear estimator relies on the fact that after averaging over many realizations of galaxy number counts while keeping realizations of lensing and optical depth fixed, the mean galaxy number count satisfy
\be
\langle\Delta_g(\bx)\rangle&=-2[1-\alpha(z)]\kappa(\bx)-\alpha(z)\tau(\bx)\,,
\ee
where brackets $\langle\,\rangle$ indicate averaging as described above. An estimator for lensing or optical depth can then be constructed by weighting observed galaxy number counts with appropriate factors [i.e. by a factor $\propto-2(1-\alpha(z)$ for lensing convergence, and by $-\alpha$ for the optical depth]. Note however that the lensing (optical-depth) estimator constructed in this way will be biased by Thomson screening (lensing), unless $\alpha\ll1$ ($\alpha\sim1$), in which case the bias will vanish. We return to the implications of reconstruction biases later in this section.  

In addition to the linear estimator, contributions to galaxy number counts that depend on density fluctuations at the quadratic order can also be used to derive estimators that reconstruct the optical-depth and lensing fluctuations.\footnote{These are: the term proportional to $\kappa(\bx)\tilde{\Delta}_g(\bx)$ and the density-deflection term defined in Eq.~\eqref{eq:lensing_realspace}, for lensing; and the term proportional to $\tilde{\Delta}_g(\bx)\tau(\bx)$ in Eq.~\eqref{eq:tau_realspace} for Thomson screening.} The underlying principle in this case is that the small-scale galaxy power spectrum (defined in Eq.~\eqref{eq:galaxy_powerspec}), which is well-observed to be statistically isotropic at leading order on large scales, becomes \textit{statistically anisotropic} in the presence of these effects, where the anisotropy is determined by the large-scale fluctuations in lensing and optical depth. Note that the word ``anisotropic'' is overloaded here: it is anisotropic as viewed by us, as statistics such as the observed number counts and power spectra vary in different positions on the sphere, but there can also be anisotropy in the sense that the correlations at a given location may not be statistically azimuthally symmetric on the sky around that point. 

Schematically, such a `quadratic' estimator of a `$X$' field (constructed from fluctuations of galaxy number density) has the form 
\be\label{eq:esimator_quad}
\hat{X}(\bL)={\scaleobj{0.8}\int\!\!}_{\bl}F_X(\bl,\bL) \Delta_g(\bl)\Delta_g(\bl-\bL)
\ee 
in Fourier space, where $\bL$ is the Fourier wave-vector satisfying $\bL\ll\bl$, and $F_X$ is a filter that is to be chosen to give an unbiased, minimum-variance reconstruction of lensing (or optical-depth) fluctuations, for example. 

The angular power-spectra of the reconstructed fields can be defined as
\begin{equation}
\langle\hat{X}(\bL)\hat{X}(\bL')\rangle=(2\pi)^2\delta^2_D(\bL+\bL')[C_L^{XX}+N_L^{XX}+\mathcal{B}_L^{XX}]\,,
\end{equation}
where $N_L^{XX}$ is the reconstruction noise of the $\hat{X}$ estimator and $C_L^{XX}$ is the angular power spectrum signal of the `$X$' field. Here, $\mathcal{B}^{XX}_L$ corresponds to the bias picked up by the estimator due to sources modifying the galaxy number count distributions \textit{other than} $X$. In our analysis, we have $X\in\{{\kappa},\tau\}$. We define our estimators and calculate the reconstruction noise spectra for lensing and optical depth in Appendix~\ref{app:estimators}. Unless stated otherwise, we use $N_L^{XX}$ when referring to the quadratic-estimator reconstruction noise, and $N_{L,\rm tot}^{XX}$ when referring to total minimum-variance reconstruction noise which we calculate by combining both the linear and quadratic estimators, as defined in Appendix~\ref{app:estimators}.

In Fig.~\ref{fig:cltt_biases} we show the optical-depth reconstruction-noise spectra from three redshift bins and LSST specifications as described before. Here, we have set the maximum multipole in the calculation of estimator-variance [see Eq.~\eqref{eq:quadnoise_tt}] to $\ell_{\rm max}=5000$ (corresponding approximately to arcminute scales). The $\ell_{\rm max}$ value sets the smallest scales we consider in the quadratic estimator (i.e. the large-$\ell$ limit of the integral in Eq.~\eqref{eq:esimator_quad}). The $x$-axes are the multipoles, $L$, corresponding to the reconstructed field. For large $L$ values and lower redshifts, we find the linear estimator improves the total reconstruction noise. The difference between $N_{L,\rm tot}^{\tau\tau}$ and $N_L^{\tau\tau}$ (gray dot-dashed and dark-red solid lines, respectively) depends strongly on the experimental specifications, including the magnification bias and the choice of $\ell_{\rm max}$, which we discuss in Appendices~\ref{app:mag_nias}~and~\ref{sec:appendix}. 

The light-red lines on these panels correspond to the lensing bias to optical-depth reconstruction. As we discussed in the previous section, the reconstruction of optical depth is biased due to lensing, since lensing magnification affects the galaxy number counts in a similar way to Thomson screening.\footnote{The contribution from the shear term defined in Eq.~\eqref{eq:lensing_realspace} also contributes as a bias to optical-depth reconstruction, albeit at a lower significance due to its different (dipolar) spatial profile.} We calculate this bias in Sec.~\ref{sec:estimators_biases}. The lensing-reconstruction noise and signal spectra are shown in Fig.~\ref{fig:clpp_biases} for redshifts and experimental specifications matching Figs.~\ref{fig:clgg_biases}~and~\ref{fig:cltt_biases}. The dependence of these results on $\alpha(z)$ and $\ell_{\rm max}$ is discussed in Appendices~\ref{app:mag_nias}~and~\ref{sec:appendix}. The orange lines in these panels correspond to the optical-depth bias on lensing reconstruction.

\begin{figure}[t!]
    \centering
    \includegraphics[width=0.49\textwidth]{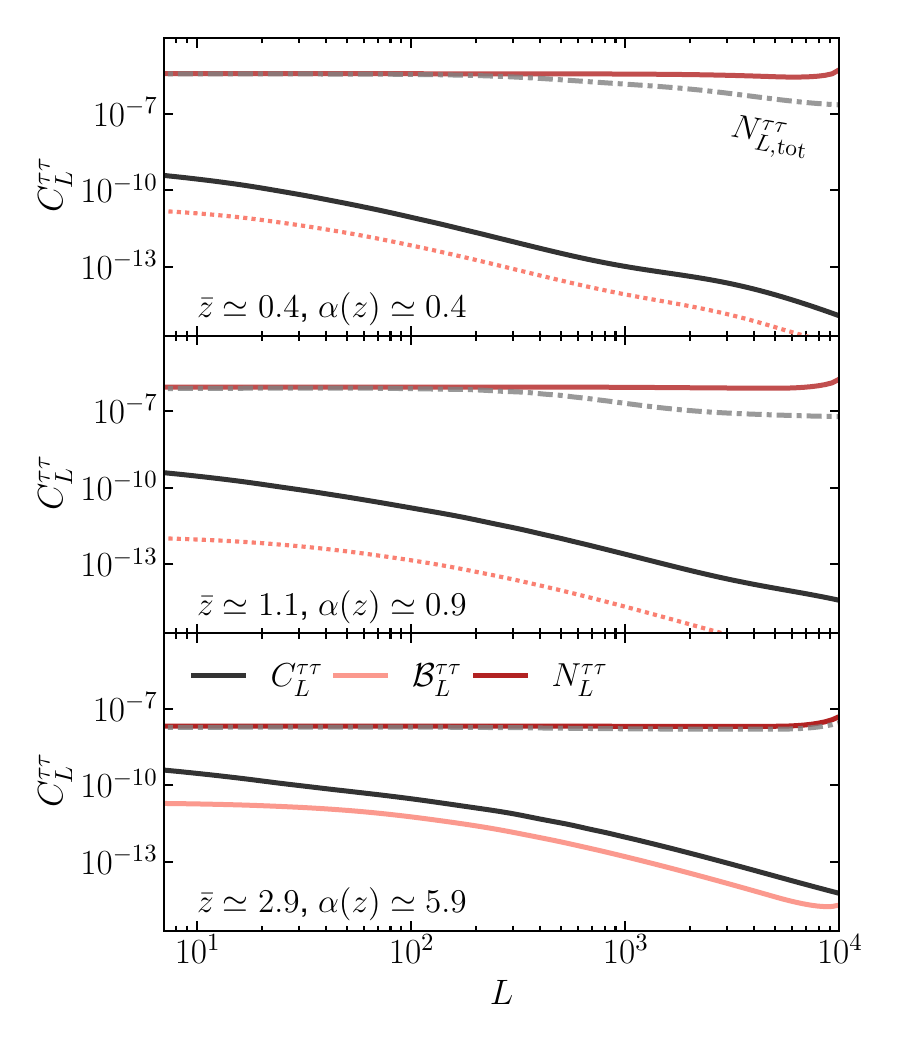}
    \vspace*{-0.7cm}
    \caption{\textit{The optical-depth reconstruction noise and signal.} Panels correspond to redshift bins as described in Fig.~\ref{fig:clgg_biases}. The solid black lines display the underlying optical-depth spectrum. The light red lines show the lensing bias on the optical-depth reconstruction. The dark red lines correspond to the optical-depth quadratic-estimator reconstruction noise. The gray dot-dashed lines correspond to the optical reconstruction noise from combination of the linear and quadratic estimators, as described in Sec.~\ref{sec:estimators}. These results depend strongly on the choice of magnification bias and the maximum multipole $\ell_{\rm max}$ considered in the quadratic-estimator integral in Eq.~\eqref{eq:quadnoise_tt} (see Appendices~\ref{app:mag_nias}~and~\ref{sec:appendix} for a detailed discussion). Here we set $\ell_{\rm max}=5000$. Lines are drawn solid and thick (thin and dotted) for positive (negative) values.}  \label{fig:cltt_biases}
    \vspace*{-0.5cm}
\end{figure}

\begin{figure}[t!]
    \centering
    \includegraphics[width=0.49\textwidth]{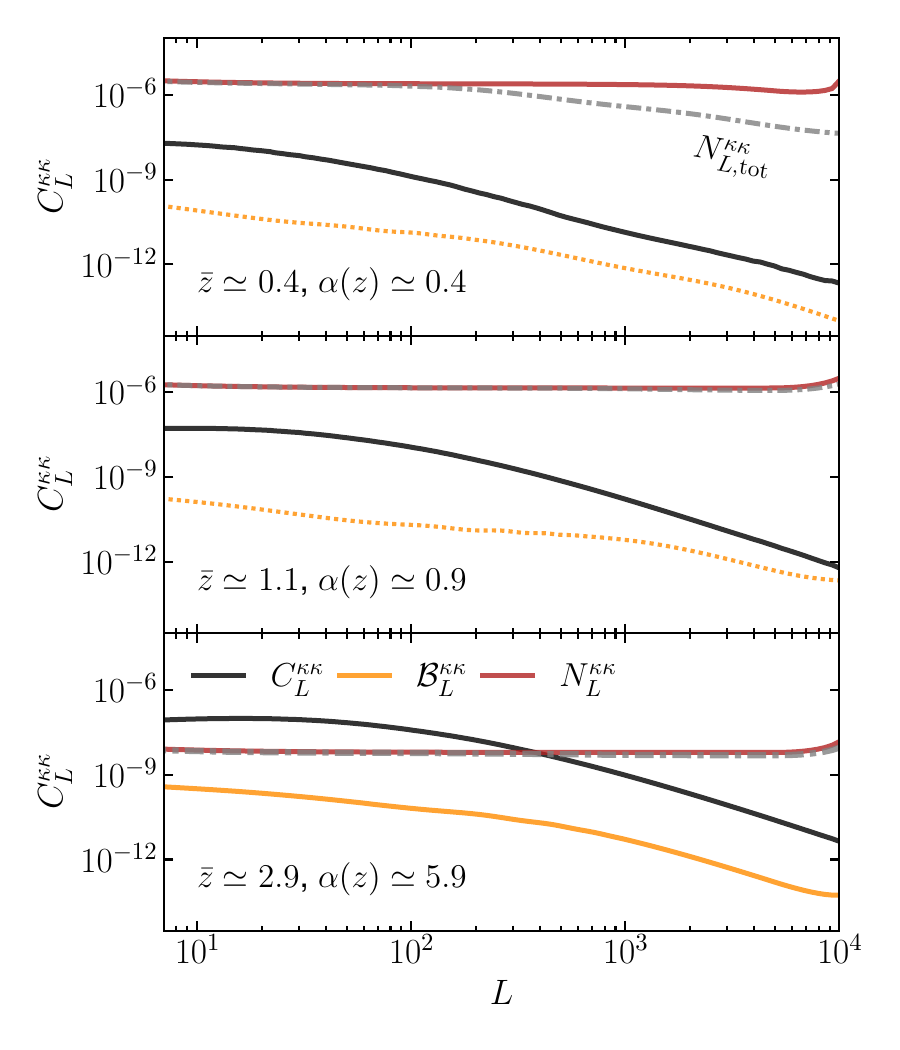}
    \vspace*{-0.7cm}
    \caption{\textit{The lensing reconstruction noise and signal}. The redshift bin and $\ell_{\rm max}$ specifications match Figs.~\ref{fig:clgg_biases}~and~\ref{fig:cltt_biases}. The orange lines correspond to the bias from optical depth discussed in Sec.~\ref{sec:estimators_biases}. The black lines correspond to the lensing signal. The red solid (gray dot-dashed) lines correspond to lensing reconstruction noise from the quadratic estimator (combined total estimator).  Lines are drawn solid and thick (thin and dotted) for positive (negative) values.}    \label{fig:clpp_biases}
    \vspace*{-0.5cm}
\end{figure}

\vspace*{-0.2cm}
\section{Results}\label{sec:results}\vspace*{-0.3cm}

\begin{table}[t!]
\centering
\begin{tabular}{ |c||c|c| } 
 \hline
 \multicolumn{3}{|c|}{Experimental Specifications ($m_{\rm lim}=25$)} \\
 \hline
 redshift $z$ & $\alpha(z)$ & $\dd n_g(z)/\dd z\,[{\rm arcmin}^{-2}]$ \\ 
 \hline
 0.26 & 0.40 & 15.4 \\% 0.37 & & 25.6  \\
 0.38 & 0.44 & 20.8 \\%  0.38 & & 38.6 \\
 0.50 & 0.49 & 22.9 \\%  0.40 & & 47.7 \\
 0.64 & 0.55 & 21.8 \\%  0.42 & & 51.6 \\
 0.79 & 0.63 & 18.6 \\%  0.46 & & 50.4 \\
 0.96 & 0.74 & 14.4 \\%  0.49 & & 45.3 \\
 1.14 & 0.90 & 10.0 \\%  0.55 & & 37.4 \\
 1.35 & 1.14 & 6.34 \\%  0.62 & & 28.4 \\
 1.58 & 1.49 & 3.57 \\%  0.71 & & 19.7 \\
 1.84 & 2.09 & 1.77 \\%  0.85 & & 12.4 \\
 2.15 & 3.23 & 0.75 \\%  1.10 & & 6.9 \\
 2.49 & 5.15 & 0.27 \\%  1.56 & & 3.4 \\
 2.90 & 5.93 & 0.08 \\%  2.10 & & 1.4 \\
 3.37 & 6.85 & 0.02 \\%  2.74 & & 0.47 \\
 3.93 & 7.94 & 0.003 \\% 3.50 &  & 0.12 \\
 4.60 & \,\,\,\,\,\,\,\,\,\,\,\,\,9.24\,\,\,\,\,\,\,\,\,\,\,\,\, & 0.0003 \\% 4.41 & & 0.02 \\
 \hline
\end{tabular}
\caption{Experimental specifications matching LSST `gold sample' with $m_{\rm lim}=25$ flux cut. We use 16 redshift bins within $z\in[0.2,5.0]$ with equal width in comoving distance (approximately 400\,Mpc). }\label{tab:lsst}
\vspace*{-0.5cm}
\end{table}

Throughout, we consider a LSST-like survey with mass cuts corresponding to $m_{\rm lim}=25$. Our choices for the magnification bias and galaxy number density are shown in Table~\ref{tab:lsst}. We calculate the quadratic and total noise spectra for lensing and optical-depth reconstruction as defined in Eqs.~(\ref{eq:quadnoise_tt}-\ref{eq:totalnoise_tt}) and Eqs.~(\ref{eq:quadnoise_pp}-\ref{eq:totalnoise_pp}). We find optical-depth reconstruction noise to be multiple-orders of magnitude greater than the signal on all scales and redshifts, while lensing reconstruction noise becomes comparable to the lensing signal at high redshifts. 

We show the sensitivity of these results on the survey specifications in Appendices~\ref{app:mag_nias}~and~\ref{sec:appendix}. In Sec.~\ref{sec:det_SNR} we calculate the detection signal-to-noise ratio (SNR) of the reconstructed optical-depth fluctuations. In Sec.~\ref{sec:fisher1} we parametrize the mean ionization fraction at a range of redshifts and forecast the measurement precision provided from optical-depth reconstruction, while Sec.~\ref{sec:fisher2} discusses the prospects of measuring parameters characterizing helium reionization.  

\vspace*{-0.2cm}
\subsection{Detection SNR}\label{sec:det_SNR}\vspace*{-0.3cm}

\begin{figure}[t!]
    \centering
    \includegraphics[width=0.49\textwidth]{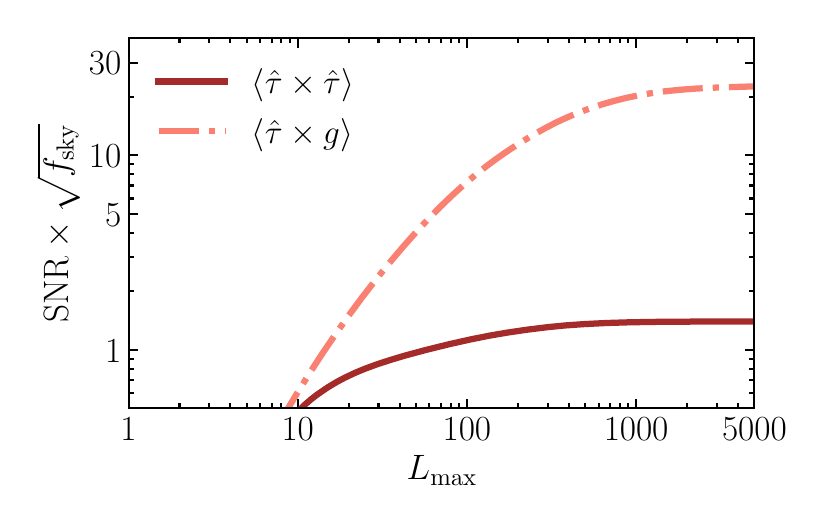}
    \vspace*{-0.7cm}
    \caption{\textit{Detection SNR of optical depth reconstruction.} The solid dark-red line corresponds to detection signal-to-noise (SNR) from auto-correlation of reconstructed optical-depth fluctuations. The dot-dashed light-red line shows the SNR from cross-correlation of the reconstructed optical-depth and galaxy density.   }\label{fig:detection_SNR_v1_lmax5K}
    \vspace*{-0.4cm}
\end{figure}

In Fig.~\ref{fig:detection_SNR_v1_lmax5K} we show the detection SNR of the optical-depth from a LSST-like survey, with maximum multipole $\ell_{\rm max}$ in Eq.~\eqref{eq:quadnoise_tt} set equal to $5000$. The solid dark-red line corresponds to the detection SNR from auto-correlations of reconstructed optical-depth fluctuations. The dot-dashed orange line correspond to detection SNR from the cross-correlation of reconstructed optical depth with the galaxy number counts.

We define the detection SNR as
\be\label{eq:SNR}
 {\rm SNR}^2\!=\!\!\!\!\!\!\!\!\sum\limits_{LL';XYWZ}\!\!\!\!\!C_L^{{X} {Y}}\boldsymbol{\mathcal{C}}^{-1}\!\!\left(\tilde{C}^{{X} {Y}}_{L},\tilde{C}^{{W} {Z}}_{L'}\right) C_{L'}^{{W}Z} \, ,
\ee
where $\boldsymbol{\mathcal{C}}(\tilde{C}^{XY}_{L},\tilde{C}^{WZ}_{L'})=  [\delta_{LL'}f_\mathrm{sky}^{-1}/({2L\!+\!1})](\tilde{C}_{L}^{XW}\tilde{C}_{L}^{YZ}\!+\!\tilde{C}_{L}^{XZ}\tilde{C}_{L}^{YW'})$ is the covariance, tilde indicates spectra including noise, and indices run over spectra at different redshift bins. We find that for the optical-depth reconstruction alone (line labelled $\langle\hat{\tau}\times\hat{\tau}\rangle$ in Fig.~\ref{fig:detection_SNR_v1_lmax5K}), detection SNR reaches $\sim1$ at $L_{\rm max}\gtrsim1000$ and $f_{\rm sky}=1$ where $f_{\rm sky}$ is the galaxy survey sky fraction (however remains below unity for $f_{\rm sky}\simeq0.4$ anticipated for LSST). Here, $L_{\rm max}$ is the maximum multipole considered in the SNR calculation in Eq.~\eqref{eq:SNR}.
The SNR from cross-correlation of reconstructed optical-depth and galaxies (the line labelled $\langle\hat{\tau}\times g\rangle$) reach significantly greater values compared to the auto-correlation. We find optical-depth detection SNR reaches $\sim13$ for $f_{\rm sky}=0.4$, for example, suggesting LSST will have sufficient statistical power to detect this signal in principle. These results depend significantly on the $\ell_{\rm max}$ value which we set to 5000 here (see Appendix~\ref{sec:appendix} for SNR results for a range of $\ell_{\rm max}$ values, as well as for lensing reconstruction for completeness). {The covariance matrix calculation contains contributions from lensing and lensing--optical-depth cross-correlation, which in principle could be mitigated using other measurements of lensing. We find removing lensing--optical-depth correlation as a source of noise improves SNR by around $\sim10$ percent.}

\vspace*{-0.2cm}
\subsection{Probing the free electron abundance}\label{sec:fisher1}\vspace*{-0.3cm}

Next, we assess the prospects to probe the mean ionization fraction $\bar{x}_e(z)$ (or free electron abundance) as a function of redshift from optical-depth reconstruction using galaxies. We define an integrated change in the mean ionization fraction $\bar{x}_e$ of the form 
\be\label{eq:mean_int_ion}
\Delta \bar{x}_e^{i}=\int_{z_{\rm min}^i}^{z_{\rm min}^i}\Delta\bar{x}_e(z)\,,
\ee 
over some redshift interval $z\in[z_{\rm min}^i,z_{\rm max}^i]$, and set equal $\Delta\bar{x}_e(z)$ to $E(z)\bar{x}_e(z)$, where $E(z)$ is a Gaussian function centered at the mean redshift of the integral, with width corresponding approximately to $(z_{\rm max}^i\!-\!z_{\rm min}^i)/2$. 

\begin{figure*}[th!]
    \centering
    \includegraphics[width=0.85\textwidth]{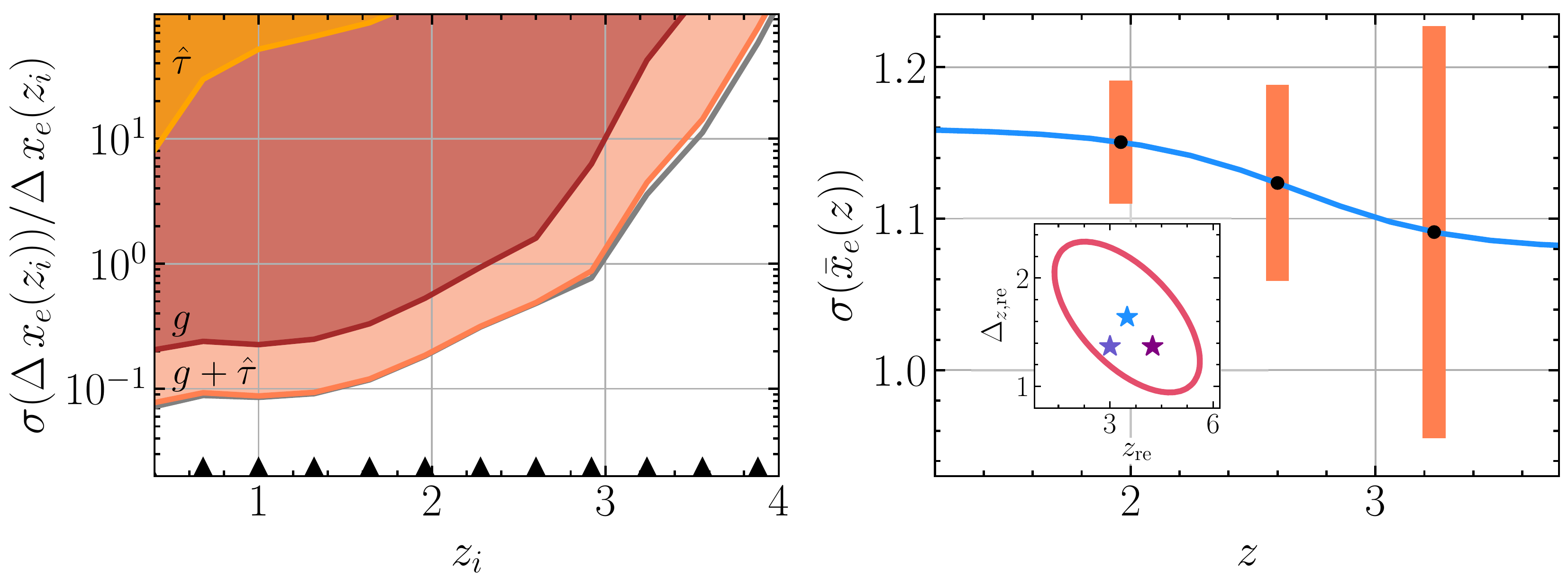}
    \vspace*{-0.4cm}
    \caption{\textit{Constraints on changes to the mean ionization fraction.}  (\textit{Left}) Panel shows the fractional constraints on the integrated change in the mean ionization fraction defined in Eq.~\eqref{eq:mean_int_ion}. The black triangle markers correspond to the centers of the redshifts intervals in Sec.~\ref{sec:fisher1}. From top, the first two shaded regions correspond to using reconstructed optical-depth and galaxy number counts in isolation. The next light-red region corresponds to using the combination of these observables. Adding reconstructed lensing only improves upon the latter marginally at higher redshifts. (\textit{Right}) The change of ionization fraction during helium reionization at higher redshifts and corresponding error from combination of reconstructed optical depth and galaxies. The inset displays the 1$\sigma$ errors on the parameters characterizing the redshift and duration of helium reionization. The star markers correspond to predicted values from three distinct helium reionization models as described in the text. These panels assume $f_{\rm sky}=0.4$.}\label{fig:results_xe}
    \vspace*{-0.5cm}
\end{figure*}

We show the fractional errors $\sigma(\Delta\bar{x}_e^{i})/\Delta \bar{x}^i_e$ on this parameter in the left panel of Fig.~\ref{fig:results_xe}, for 14 equal-redshift intervals within $z\in[0.2,5.0]$, which we calculate by defining an information matrix in the form
\be\label{eq:fisher}
\mathcal{F}_{ab}=\!\!\!\!\!\!\sum\limits_{LL';XYWZ}\!\!\!\!\frac{\partial C_L^{X Y}}{\partial {\pi_a}}\boldsymbol{\mathcal{C}}^{-1} \left(\tilde{C}^{X W}_{L},\tilde{C}^{Y Z}_{L'}\right)\frac{\partial C_{L'}^{W {Z}}}{\partial {\pi_b}}\,,
\ee
where $\pi_{a}$ vary over the parameter array consisting of parameters characterizing the ionization fraction $\Delta \bar{x}^i_e$ ($i\in\{1,\ldots,14\}$) and galaxy bias parameters $b_g^j$ for each redshift bin ($j\in\{1,\ldots,16\}$). In addition to these, we also include parameters to capture the modelling uncertainty of the magnification bias, which we define with $\alpha^{\rm obs}(\bar{z}_j)=b_\alpha^j \alpha^{\rm true}(\bar{z}_j)$, with fiducial values set to unity.\footnote{In principle, modelling uncertainties of the underlying small-scale galaxy power spectra could also bias the reconstruction of cosmological fluctuations from small-scale statistical anisotropies. For the $\ell_{\rm max}=5000$ choice we make here, we assume this bias will be small and reasonably-well understood.}

We find reconstructed optical-depth fluctuations provide $\sim10$ percent fractional constraints on the integrated change of ionization fraction defined in Eq.~\eqref{eq:mean_int_ion} at redshifts below $z\sim1.5$ for $f_{\rm sky}=0.4$. Constraints worsen significantly with increasing redshift. The galaxy spectrum alone also constrains ionisation fraction through the Thomson screening terms given in Eq.~\eqref{eq:galaxy_total}, albeit a factor $\sim3$ weaker compared as when combined with reconstructed optical depth. The reconstruction provides weak constraints when considered in isolation. Adding lensing reconstruction on top of galaxy and optical depth improves the constraints only marginally (gray colored region extending below region labelled `$g+\hat{\tau}$'). We find marginalizing over galaxy bias and $b_\alpha^j$ do not effect the fractional constraints on $\Delta x_e(z)$ beyond $\sim0.1$ percent. We leave evaluating how various deviations from the standard cosmological model of mean ionization fraction (or electron abundance) map onto these constraints (briefly motivated in Sec.~\ref{sec:Intro}) to upcoming work. Next, we discuss the prospects for probing helium reionization.

\vspace*{-0.2cm}
\subsection{Probing helium reionization}\label{sec:fisher2}\vspace*{-0.3cm}

The right panel in Fig.~\ref{fig:results_xe} corresponds to selecting 3 redshift intervals of $\Delta\bar{x}_e(z)$ around the epoch of helium reionization $2<z<4$. Reducing the number of $\Delta \bar{x}_e^i$ parameters from the parameter vector significantly improves the constraints, since $\Delta \bar{x}_e^i$ are significantly degenerate with each other. Note that here we set $f_{\rm sky}=0.4$. The orange error bars correspond to the combination of galaxy, reconstructed optical-depth and lensing fluctuations. (Constraints are similarly dominated by the first two.) The blue line is the prediction for the ionization fraction given some fiducial choices of helium reionization parameters defined with a hyperbolic tangent
\be\label{eq:mean_reio}
\overline{x}_e(z)\!=\!\frac{1}{2}\!\left[2\!+\!\Delta \bar{x}\!-\!\Delta \bar{x}\tanh{\left(\frac{y(z_{\rm re})\!-\!y(z)}{\Delta_y}\right)}\right]\!,\,\,\,\,\,
\ee
as commonly done in the CMB literature~\citep[e.g.][]{Lewis:2008wr}. Here, $y(z)=(1+z)^{3/2}$, $\Delta \bar{x}_{\rm He}$ determines the total change in the mean ionization fraction during helium reionization, $z_{\rm re}$ is the redshift half-way-through the helium reionization and $\Delta_y$ parameterize the duration of the transition. Note we trade $\Delta_y$ with $\Delta_{z,\rm re}$, which we define as the duration in redshift of the central 50$\%$ change in ionization fraction. We set the fiducial parameters $\{\Delta \bar{x}_{\rm He},z_{\rm re},\Delta_{z,\rm re}\}$ to $\{$0.08,2.5,0.5$\}$ for the blue line. 

The inset contour plot in the left panel of Fig.~\ref{fig:results_xe} shows the $1\sigma$ constraint on $\Delta_{z,\rm re}$ and $z_{\rm re}$ from a similar analysis described in Sec.~\ref{sec:fisher1} using helium reionization parameters rather than $\Delta \bar{x}_e^i$. We set the fiducial values for ($z_{\rm re}$, $\Delta_{z,\rm re}$) to (3.34, 0.8), which reproduce quasar properties measured in Refs.~\citep{2013ApJ...773...14R, Masters2012, 2013ApJ...768..105M, 2012MNRAS.424..933W, 2015MNRAS.449.4204L}. The markers shown in that contour plot corresponds to fitted values from assuming quasar abundance reduced by a factor of 2 (green marker) and a uniform UV background rather than explicit quasar sources (purple marker). 

We find the measurement precision on helium reionization parameters using the method described in this work is modest, likely insufficient to distinguish between different helium reionization models unambiguously at the precision of LSST. Note however that with $1\sigma$ errors on the time (duration) of reionization around $2\,(1)$ in redshift, it is suggestive that LSST with $m_{\rm lim}=25$ can in principle detect evidence of helium reionization from Thomson screening. 

\vspace*{-0.2cm}
\section{Discussion}\label{sec:discussion}\vspace*{-0.3cm}

Precise measurements of galaxy number counts will open the window into using galaxy surveys as a back-light for large-scale structure, similar to the CMB. The prospect of measuring the effect of lensing on the galaxy number count distribution has been studied in the literature~\citep[e.g.][]{2002ApJ...580L...3J, 2005ApJ...633..589S, Hoekstra:2008db, 2009A&A...507..683H, 2013MNRAS.429.3230H, 2011PhRvD..84j3004V, 2012ApJ...744L..22S, 2019MNRAS.482..785S, 2021MNRAS.504.1452V,Nistane:2022xuz,Wenzl:2023wsd,Ma:2023kyc} and we expand these ideas to probe cosmological electron fluctuations via Thomson. In this paper we showed that upcoming galaxy surveys such as LSST can detect the effect of patchy Thomson screening, providing an alternative probe of the abundance of electrons throughout cosmic history. 

We performed forecasts for a LSST-like galaxy survey, with specifications given in Table~\ref{tab:lsst}, assuming a flux cut satisfying $m_{\rm lim}<25$. As our results depend significantly on these choices,  further study on optimizing galaxy survey specifications to maximize the prospects for lensing and optical-depth reconstruction could be fruitful. 

An assumption that we made throughout was that the galaxy density field could be treated as a Gaussian random field. A cursory look at any galaxy surveys reveals that they are not in fact Gaussian random fields, with substantial intrinsic non-Gaussianity that will act as additional noise if neglected for these estimators. This has been extensively studied for the case of intensity mapping (e.g., \cite{Foreman:2018lci}), and more study will be required for the case of galaxy number counts. It may be possible to take advantage of tweaks to  estimators that may make them less sensitive to intrinsic non-Gaussianity \citep{namikawa2013bias,2019PhRvL.122r1301S}. It also may be possible to include a model for the gravitationally induced non-Gaussianity as part of the statistically isotropic signal that is being affected by the intervening large scale structure. 

Another promising route to boosting the detection prospects of lensing and optical-depth reconstruction following the methods we discussed in our paper is considering other observables for cross-correlation. These include galaxy-lensing shear, for example, which we did not consider in this study. The cross-correlations of reconstructed lensing fluctuations with the galaxy shear has recently been shown to be a promising way to increase the detection SNR of galaxy lensing~\citep{Buncher:2024hyt}. In an upcoming work we will demonstrate the benefit from including other tracers of large-scale structure including lensing shear in our analysis, as well as point out the limiting factor these distortions can play on the measurement precision of lensing-shear--galaxy cross-correlation; one of the major science programs for galaxy surveys in the near future. 

Most notably, we have not considered the effect of inter-galactic dust on the galaxy number counts~\citep[e.g.][]{2010MNRAS.405.1025M,Menard:2007rd}. Dust is well-understood to be present around galaxies ~\citep[e.g.][]{2010ApJ...716..712A,2012ApJ...746...85S,2015ApJ...813....7P}, {with an effect that can exceed} that of free electrons~\citep[see e.g.][for a recent analysis]{Rubin:2023ovl}. It is not clear how well dust can survive the harsh UV background well outside galaxies, but even without large amounts of intergalactic dust there could be substantial effects.  Dust leads to a screening similar to Thomson screening, but with a characteristic frequency-dependent screening (``reddening''). Recent studies show that intergalactic dust could contribute significantly to the measurement uncertainties of galaxies~\citep{Rubin:2023ovl}. The effects of dust (i.e. dust extinction) could be `corrected' by modelling and exploiting the frequency-dependence of this signal (i.e dust correction)~\citep[e.g.][]{Graham:2008hn,2012A&A...539A..31C,Schmidt:2014jja} or measured jointly with galaxy density, for example~\citep[e.g.][]{Bravo:2021att}. It is also possible to use the dust-induced extinction as also a cosmological signal~\citep[e.g.][]{2010MNRAS.405.1025M}, just as we are doing here with the frequency-independent Thomson screening. 

We also omitted assessing the prospects of cross-correlating reconstructed optical depth and lensing with other tomographic measurements of cosmological fluctuations, such as CMB lensing reconstruction~\citep{Hu:2001kj}, radial velocities from kSZ~\citep{Cayuso:2021ljq}, transverse velocities from moving-lens effects~\citep{Hotinli:2018yyc,Hotinli:2020ntd,Hotinli:2021hih}, polarized SZ effects~\citep{Hotinli:2022wbk,Lee:2022udm}, as well as optical-depth and velocity-square reconstructions from CMB~\citep{Dvorkin:2008tf,Smith:2016lnt,Schutt:2024zxe}, which may boost the prospects of probing the ionization fraction as a function of redshift, once analysed jointly. The small-scale patchy screening of CMB photons was detected recently in cross-correlation with galaxies~\citep{ACT:2024rue}.

{Finally, note that in this paper we have treated the reconstruction biases as contribution to the optical depth (and lensing) reconstruction covariance. In doing so we take into account the increase in the correlation of the reconstructed fields at different redshift bins, for example, while we omit accounting for the degeneracy between optical depth and lensing fields. In practice, these biases will impair our prospects of unambiguously reconstructing lensing and optical-depth fluctuations from galaxy surveys, as they can confuse each other.} In principle one can use bias-hardened estimators~\citep[e.g.][]{2013MNRAS.431..609N} by taking only the Fourier modes of galaxy density that have orthogonal contributions from Thomson screening and lensing, for example. Since such estimators use only a subset of the available small-scale Fourier modes, they consequentially have greater reconstruction noise. {Alternatively, one could use the different dependence on the slope of the number counts $\alpha(z)$ to separate optical depth from lensing by doing this for a range of cuts that all have different magnification bias.} We leave developing better unbiased estimators for galaxy lensing and optical-depth to future work.

Measurements of free electron fluctuations are a promising probe of various deviations from LCDM and the standard model of particle physics, as well as astrophysical phenomena such as helium reionization. The number of cosmological probes of electrons is limited, making new methods such as the one introduced in this paper extremely valuable. With an array of large-scale structure surveys coming online, the future of cosmology is promising. We argue that galaxies can serve as a cosmological back-light on intervening structure on top of the well-studied signatures such as clustering, enhancing the cosmological information accessible from these surveys.  

\vspace*{-0.2cm}
\section{Acknowledgements}
\vspace*{-0.2cm}

We thank Brandon Buncher, Simone Ferraro, Matthew Johnson, Marc Kamionkowski, Mathew Madhavacheril, Brice Menard, Kendrick Smith, for useful discussions. SCH is supported by the P.~J.~E.~Peebles Fellowship at Perimeter Institute for Theoretical Physics. This work was supported by Brand and Monica Fortner and the Canadian Institute for Advanced Research. This research was supported in part by Perimeter Institute for Theoretical Physics. Research at Perimeter Institute is supported by the Government of Canada through the Department of Innovation, Science and Economic Development Canada and by the Province of Ontario through the Ministry of Research, Innovation and Science. This work was performed in part at Aspen Center for Physics, which is supported by National Science Foundation grant PHY-2210452. This work was in part carried out at the Advanced Research Computing at Hopkins (ARCH) core facility  (rockfish.jhu.edu), which is supported by the National Science Foundation (NSF) grant number OAC1920103.

\appendix

\vspace*{-0.2cm}
\section{Galaxy number counts}\label{sec:galaxy_fourier}
\vspace*{-0.2cm}

In Fourier $\bl$-space and using the flat-sky approximation, the number-density field (defined in Eq.~\eqref{eq:galaxy_realspace}) takes form
\begin{equation}
\begin{split}
\Delta_g(\bl,z)\!=\!\tilde{\Delta}_g(\bl,z)\!+\!\delta{\Delta}^{\rm Lens}_g(\bl,z)\!+\!\delta{\Delta}^{\rm Thomson}_g(\bl,z)
\end{split}
\end{equation}
with
\begin{equation}
    \begin{split}
    \delta&\tilde{\Delta}^{\rm Lens}_g(\bl,z)\\
    &={\!r(z){\kappa(\bl,z)}}+{\!{\scaleobj{0.8}\int\!\!}_{\bl'}\!\! \tilde{\Delta}_g(\bl',z) \phi(\bl\!-\!\bl',z)P(\bl,\bl',z)}\,,
    \end{split}
\end{equation}
where we defined
\be 
P(\bl,\bl',z)\equiv\!(1\!-\!\alpha(z))|\bl-\bl'|^2+\bl'\cdot(\bl-\bl')\,,
\ee 
and {$r(z)\equiv2(1-\alpha(z))$}, while the Thomson scattering term satisfies
\begin{equation}
    \begin{split}
        \delta&\Delta_g^{\rm Thomson}(\bl,z)\\
        &=-\alpha(z)\, \tau(\bl,z)-\!\alpha(z)\!\!{\scaleobj{0.8}\int\!\!}_{\bl'}\! \!\tilde{\Delta}_g(\bl',z) \tau(\bl-\bl',z)\,.
    \end{split}
\end{equation}
The galaxy power spectrum can be calculated as
\be
\langle\Delta_g(\bl,z)\Delta_g(\bl',z')\rangle=(2\pi)^2\delta^2(\bl\!+\!\bl') C_\ell^{\Delta_g\Delta_g(z,z')}\,,
\ee 
which satisfy
\begin{equation}
\begin{split}
{C}_{\ell}^{\Delta_g\Delta_g(z,z')}&\simeq  \, C_\ell^{\tilde{\Delta}_g\tilde{\Delta}_g(z,z')}
\\&+r(z)r(z'){C_\ell^{\kappa\kappa(z,z')}}+ \alpha(z)\alpha(z')C_\ell^{\tau\tau(z,z')}\\
&{-r(z')C_\ell^{\Delta_g\kappa(z,z')}
 - r(z)C_\ell^{\Delta_g\kappa(z',z)}}  \\
&- \alpha(z') C_\ell^{\Delta_g\tau(z,z')}-\alpha(z) C_\ell^{\Delta_g\tau(z',z)} \\
&+ r(z)\alpha(z') {C_\ell^{\kappa\tau(z,z')}+r(z')\alpha(z) C_\ell^{\kappa\tau(z',z)}\,.}
\end{split}\label{eq:galaxy_total}
\end{equation}
We show the galaxy power spectrum together with lensing and Thomson screening contributions in Fig.~\ref{fig:clgg_biases} for the three redshift bins and magnification bias $\alpha(z)$ matching LSST `gold sample', as defined in Sec.~\ref{sec:Num_count_distortions}. We discuss the dependence of these results to the choice of magnification bias in Appendix~\ref{app:mag_bias_gal}.

\vspace*{-0.2cm}
\section{Estimators}\label{app:estimators}
\vspace*{-0.2cm}

The mean and the power spectrum of galaxy number counts, $\tilde{\Delta}_g$, averaged over many realisations fixing $\tau$ and $\phi$, satisfy
\begin{equation}
\label{eq:mean_g}
\langle\Delta_g(\bl,z)\rangle={2(1-\alpha(z))\kappa(\bl,z)}{-\alpha(z)\tau(\bl,z)}\,,
\end{equation}
and
\begin{equation}
\begin{split}\label{eq:pows_g}
\langle\Delta_g(\bl,z)\Delta_g(\bl',&z)\rangle 
\\=(2\pi)^2 \delta^2_D&(\bl\!+\!\bl') C_\ell^{\tilde{\Delta}_g \tilde{\Delta}_g}\\ &+G_\kappa(\bl,\bl+\bl',z) \kappa(\bl\!+\!\bl')\\
&+f_\tau(\bl,\bl-\bL,z) \tau(\bl\!+\!\bl') + \!\mathcal{O}(2)\,, 
\end{split}
\end{equation}
where
\begin{equation}
\begin{split}
G_\kappa&(\bl,\bl',z)=~r(z) \big(C_{|\bl'-\bl|}^{\tilde{\Delta}_g\tilde{\Delta}_g}+C_{\ell}^{\tilde{\Delta}_g \tilde{\Delta}_g}\big)\\
&-2\big[(\bl'-\bl)\cdot\bl'\,C_{|\bl'-\bl|}^{\tilde{\Delta}_g \tilde{\Delta}_g}+(\bl \cdot\bl')C_{\ell}^{\tilde{\Delta}_g \tilde{\Delta}_g}\big]/{\ell'}^2\,,
\end{split}
\end{equation}
and
\begin{equation}
f_\tau(\bl,\bl-\bL,z)\equiv\alpha(z)\big(C_{|\bl-\bL|}^{\tilde{\Delta}_g\, \tilde{\Delta}_g(z)}
\!+\!C_{\ell}^{\tilde{\Delta}_g\, \tilde{\Delta}_g(z)}\big)\,.
\end{equation} 
Above Eqs.~\eqref{eq:mean_g}~and~\eqref{eq:pows_g} demonstrate the statistical anisotropies induced by fluctuations of optical depth and lensing. Next, we define estimators that aim to reconstruct these from the anisotropy they induce on galaxy density fluctuations. 

\vspace*{-0.2cm}
\subsection{Patchy optical depth}
\vspace*{-0.2cm}

We define the estimator for optical-depth field as 
\begin{equation}
\begin{split}
&\hat{\tau}(\bL,z)\\
&\!=\!\underbrace{(1\!-\!A_\tau(L,z))\frac{\Delta_g(\bL,z)}{g_\tau(z)}}_{\rm linear}\!+\!\underbrace{A_\tau(L,z)N_\tau(L,z)\,I_\tau(\bL,z)}_{\rm quadratic},
\end{split}
\end{equation}
where terms labelled `linear' and `quadratic' correspond to two estimators that use the effects given in Eqs.~\eqref{eq:mean_g}~and~\eqref{eq:pows_g}, respectively, $g_\tau(z)\equiv-\alpha(z)$,
\begin{equation}
I_\tau(\bL,z)\equiv{\scaleobj{0.8}\int\!\!}_{\bl}\!F_\tau(\bl,\bL\!-\!\bl,z)\Delta_g(\bl,z)\Delta_g(\bL\!-\!\bl,z)\,,
\end{equation}
and the filter $F_\tau(\bl,\bL-\bl,z)$ and normalization $N_\tau(L,z)$ satisfy an unbiased, minimum-variance $\tau$ reconstruction (in the absence of other sources of anisotropy). The filter can be found to satisfy
\begin{equation}
F_\tau(\bl,\bL-\bl,z)\equiv \frac{f_\tau(\bl,\bL-\bl,z)}{2C_{\ell}^{\Delta_g\,\Delta_g(z)}C_{|\bl-\bL|}^{\Delta_g\,\Delta_g(z)}}\,.
\end{equation}
The quadratic-estimator variance satisfies
\begin{equation}
\label{eq:quadnoise_tt}
\frac{1}{N_\tau(L,z)}\!\equiv\!{\scaleobj{0.8}\int\!\!}_{\bl'}f_\tau(\bl,\bL-\bl,z)F_\tau(\bl,\bL-\bl,z)\,,
\end{equation}
and
\begin{equation}
A_\tau(L,z)\!=\!\frac{C_L^{\Delta_g\,\Delta_g(z)}}{g_\tau(z)^2N_\tau(L,z)+C_L^{\Delta_g\,\Delta_g(z)}}\,.
\end{equation}
The total reconstruction noise then satisfies
\begin{equation}
\label{eq:totalnoise_tt}
N^{\tau\tau}_{\rm tot}(L,z)\!=\!\frac{C_L^{\Delta_g\,\Delta_g(z)}N_\tau(L,z)}{g_\tau(z)^2N_\tau(L,z)+C_L^{\Delta_g\,\Delta_g(z)}}\,.
\end{equation}

\vspace*{-0.2cm}
\subsection{Weak lensing}
\vspace*{-0.2cm}

Similar to optical depth, the lensing estimator has the form
\be
\begin{split}
&\hat{\kappa}(\bL,z)\\
&=(1\!-\!A_\kappa(L,z))\frac{\Delta_g(\bL,z)}{r(z)}\!+\!A_\kappa(L,z)N_\kappa(L,z)I_\kappa(\bL,z)\,,
\end{split}\non
\ee
where
\begin{equation}
I_\kappa(\bL,z)\equiv\!{\scaleobj{0.8}\int\!\!}_{\bl'}\!F_\kappa(\bL,\bL\!-\!\bl,z)\Delta_g(\bl,z)\Delta_g(\bL\!-\!\bl,z)\,,
\end{equation}
and
\begin{equation} 
F_\kappa(\bL,\bL-\bl,z)=\frac{f_\kappa(\bl,\bL-\bl,z)}{2C_\ell^{\Delta_g\Delta_g(z)}C_{|\bL-\bl|}^{\Delta_g\Delta_g(z)}}\,.
\end{equation}
where
\begin{equation}\label{eq:quadnoise_pp}
\begin{split}
&f_\kappa(\bl,\bL-\bl,z)\\
=&\,r(z)(C_\ell^{\tilde{\Delta}_g\tilde{\Delta}_g(z)}+C_{|\bL-\bl|}^{\tilde{\Delta}_g\tilde{\Delta}_g(z)}) \\&\!-\!2\big[\bl\cdot\bL\, C_{\ell}^{\tilde{\Delta}_g \tilde{\Delta}_g(z)}
\!+(\bL\!-\!\bl) \cdot \bL\, C_{|\bL-\bl|}^{\tilde{\Delta}_g\tilde{\Delta}_g(z)}\,\big]/L^2.
\end{split}
\end{equation}
The estimator variance satisfies 
\be 
\frac{1}{N_\kappa(L,z)}\!=\!{\scaleobj{0.8}\int\!\!}_{\bl'}f_\kappa(\bl,\bL-\bl,z)F_\kappa(\bl,\bL-\bl,z)\,,
\ee
and
\be
A_\kappa(L,z)=\frac{C_L^{\Delta_g\Delta_g(z)}}{r(z)^2N_\kappa(L,z)+C_L^{\Delta_g\Delta_g(z)}}\,.
\ee

The total lensing-reconstruction noise satisfies
\begin{equation}\label{eq:totalnoise_pp}
N^{\kappa\kappa}_{\rm tot}(L,z)=\frac{C_L^{\Delta_g\,\Delta_g(z)}N_\kappa(L,z)}{r(z)^2N_\kappa(L,z)+C_L^{\Delta_g\,\Delta_g(z)}}\,.
\end{equation}

\vspace*{-0.2cm}
\subsection{Biases}\label{sec:estimators_biases}
\vspace*{-0.2cm}

Weak gravitational lensing biases the optical depth reconstruction as

\begin{equation}\label{eq:bias1}
\begin{split}
{b}_\tau(\bL,z)=-(1-&A_\tau(L,z))\frac{{r(z)\kappa(\bL,z)}}{g_\tau(z)}
\\ &+ \kappa(\bL,z)A_\tau(L,z)N_\tau(L,z){\beta(\bL,z)}\,,
\end{split}
\end{equation}
where
\be
\beta(\bL,z)\!=\!{\scaleobj{0.8}\int\!\!}_{\bl'} \!F_\kappa(\bl,\bL-\bl,z){f_\tau(\bl,\bL,z)}\,.
\ee
Similarly, the bias on the lensing estimator from the optical-depth fluctuations is
\begin{equation}\label{eq:bias2}
\begin{split}
b_\kappa(\bL,z)=-(1-&A_\kappa(L,z))\frac{g_\tau(z)\tau(\bL,z)}{r(z)}\\
&+A_\kappa(L,z)N_\kappa(L,z)\,{\tau(\bL,z)}\beta(\bL,z)\,.
\end{split}
\end{equation}
We define the bias spectra as 
\begin{equation}
\langle b_X(\bl,z)b_X(\bl',z)\rangle=(2\pi)^2\delta^2(\bl\!+\!\bl') \mathcal{B}_\ell^{XX(z)}\,,
\end{equation}
where $X\in\{\tau,{\kappa}\}$.

The contribution of these biases to the reconstruction of optical-depth and lensing  fluctuations, labelled $\mathcal{B}_\ell^{\tau\tau}$ and {$\mathcal{B}_\ell^{\kappa\kappa}$}, are shown in Figs.~\ref{fig:cltt_biases}~and~\ref{fig:clpp_biases} with solid and thick (dotted and thin) lines where values are positive (negative). For lensing reconstruction, we find the optical-depth bias (light red lines shown in Fig.~\ref{fig:clpp_biases}) remain over an order-of-magnitude smaller than the lensing power spectrum {$C_\ell^{\kappa\kappa}$}. For the optical-depth reconstruction shown in Fig.~\ref{fig:cltt_biases}, however, the effect of lensing is more significant on all scales. These result also depend significantly on the value of magnification bias $\alpha(z)$, which we discuss in Appendix~\ref{app:mag_bias_recsig}.

\begin{figure*}[t!]
    \centering
    \includegraphics[width=0.96\textwidth]{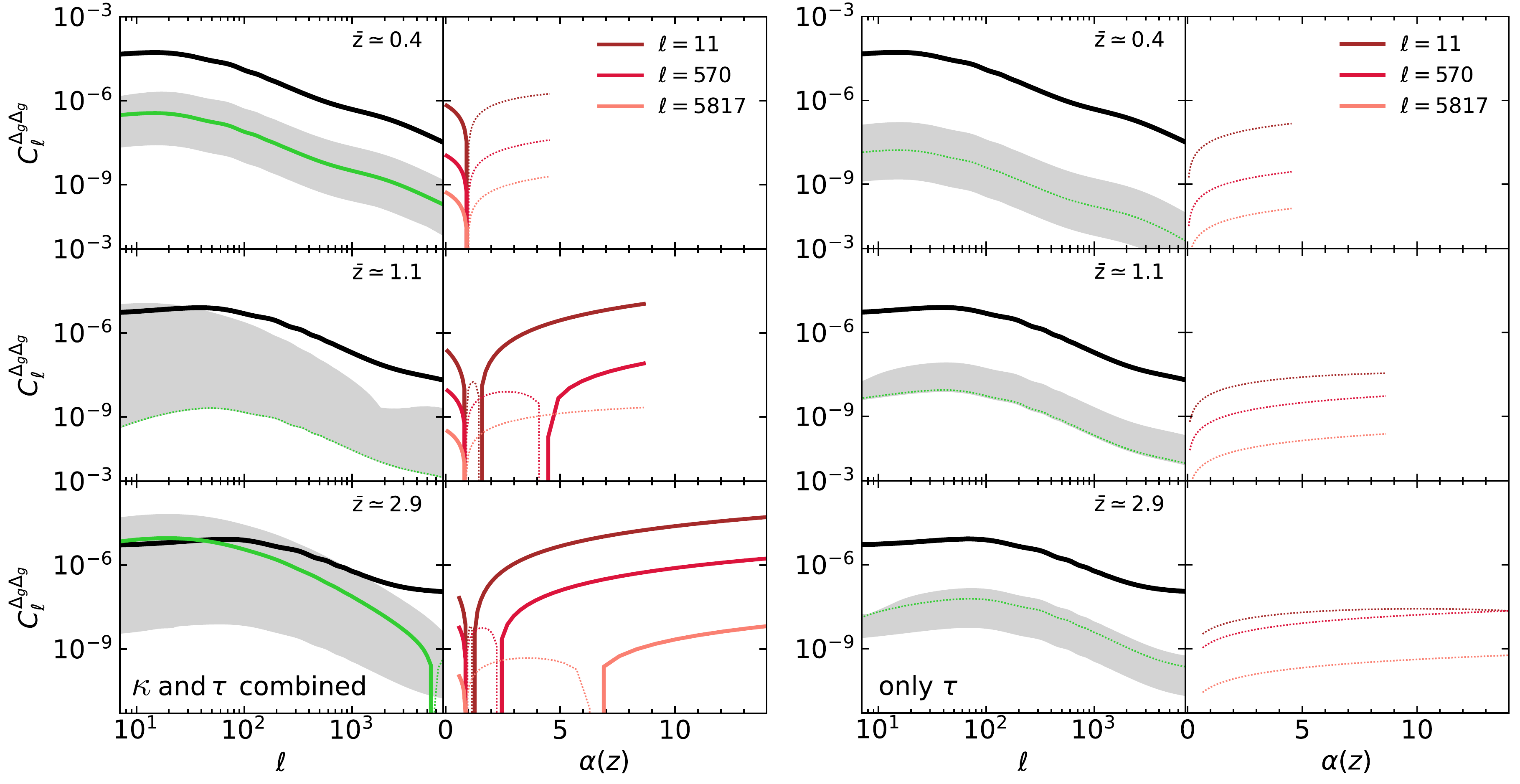}
    \vspace*{-0.35cm}
    \caption{(\textit{Left set of panels}) \textit{The galaxy number count spectra and total bias from lensing and optical depth effects combined.} The black solid lines correspond to the galaxy number count power spectra in the absence of lensing and Thomson screening. The gray-shaded regions extend between the minimum and maximum bias spectra values, calculated by setting magnification bias to vary between $[0.04,4.3]$, $[0.09,9.0]$ and $[0.6,15]$ for top, middle and bottom rows, respectively. The green lines correspond to results presented in previous sections for LSST survey `gold sample' specifications. Throughout we use solid tick (dotted thin) line style to indicate positive (negative) values. The dependence of the bias on $\alpha(z)$ is highly non-trivial for middle and bottom panels. The right column of panels show how the bias spectra amplitudes for $\ell$ values set to $\{11,570,5617\}$ vary with $\alpha(z)$. For the lowest redshift bin, the total bias is positive for $\alpha(z)<1$ (negative for $\alpha(z)>1)$. For the middle and higher redshift bins, the bias changes sign again at larger $\alpha$ values due to lensing shear term. Note that we find $\alpha(z)$ dependence of the total bias ({$\kappa$} and $\tau$ combined) is dominated by lensing for all redshift and multipoles, and the left set of panels match (within $<10$ percent) with the contribution from lensing alone (i.e. lines labelled {`only $\kappa$'} in Fig.~\ref{fig:clgg_biases}). Hence we have chosen to omit showing the contribution to galaxy spectrum from lensing alone. (\textit{Right set of panels}) \textit{The bias contribution to the galaxy spectra from Thomson screening.} Shaded regions span between the maximum and minimum values the bias takes within the same range of $\alpha(z)$ values of the left set of panels. The optical-depth bias is negative at all redshifts and multipole values we consider. The dependence of the bias to $\alpha(z)$ (right column of panels) is simpler compared to lensing due to absence of shear.}
    \label{fig:alpha_analysis2}
    \vspace*{-0.4cm}
\end{figure*}

\vspace*{-0.2cm}
\section{Dependence on $\alpha(z)$}
\label{app:mag_nias}
\vspace*{-0.2cm}

Our results throughout this paper depend strongly on the value of magnification bias $\alpha(z)$.

\begin{figure*}[t!]
    \centering
    \includegraphics[width=0.95\textwidth]{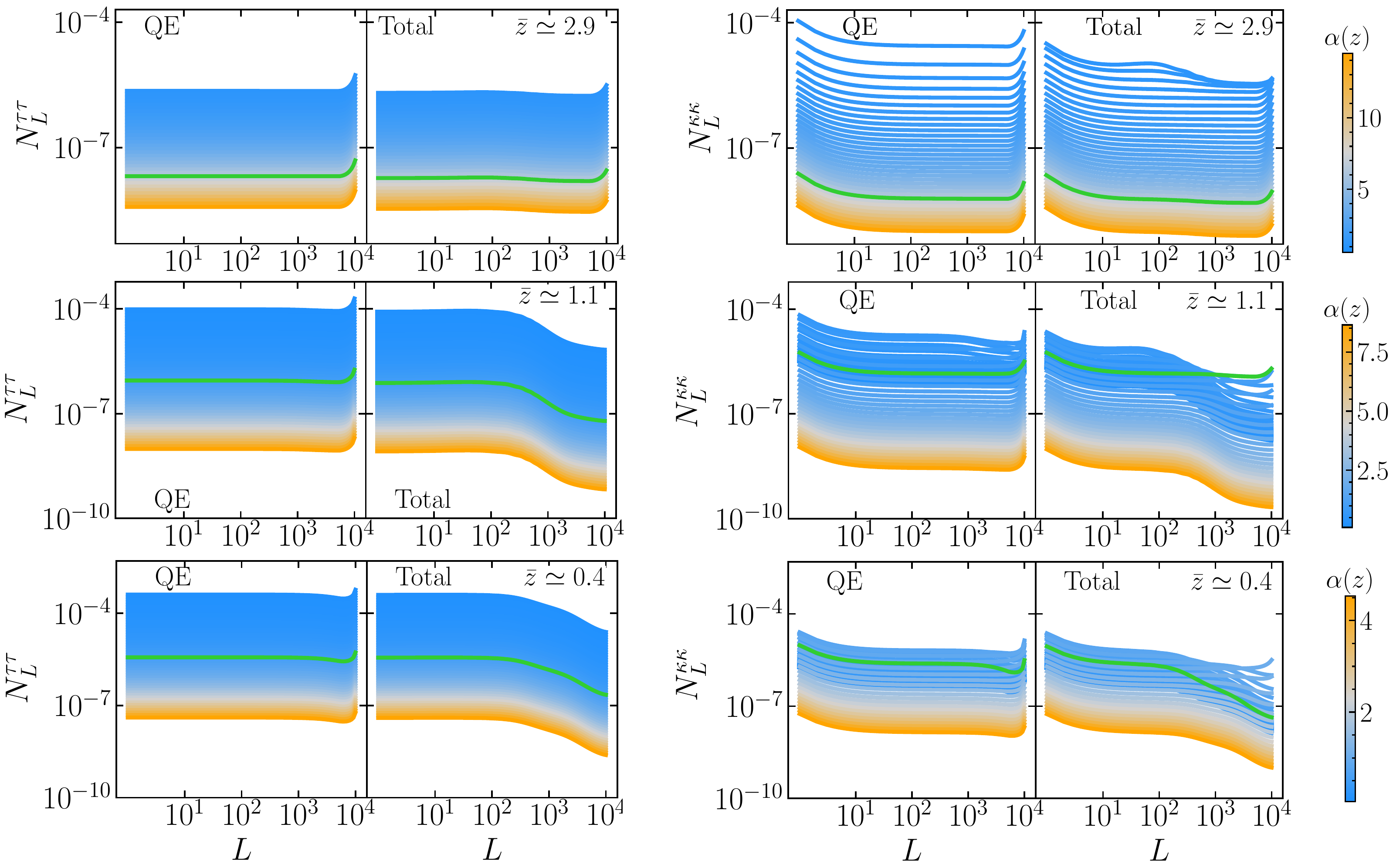}
    \vspace*{-0.5cm}
    \caption{\textit{The lensing and optical-depth reconstruction noise spectra for a range of magnification bias values}. The rows and $\alpha(z)$ ranges match Fig.~\ref{fig:alpha_analysis2}. The left (right) columns in each set of $3\times2$ panels correspond to quadratic-estimator (total) reconstruction noise labelled QE (Total). The left (right) set of panels correspond to optical-depth (lensing) reconstruction. Panes share the same $y$- and $x$-axes range. The line colors correspond to different magnification bias values given by the three rows of color-bars on the right of each row. The green lines correspond to the magnification bias choice used in previous sections, matching LSST `gold sample'.
    }\label{fig:alpha_analysis}
    \vspace*{-0.3cm}
\end{figure*}

\begin{figure*}[t!]
    \centering
    \includegraphics[width=0.96\textwidth]{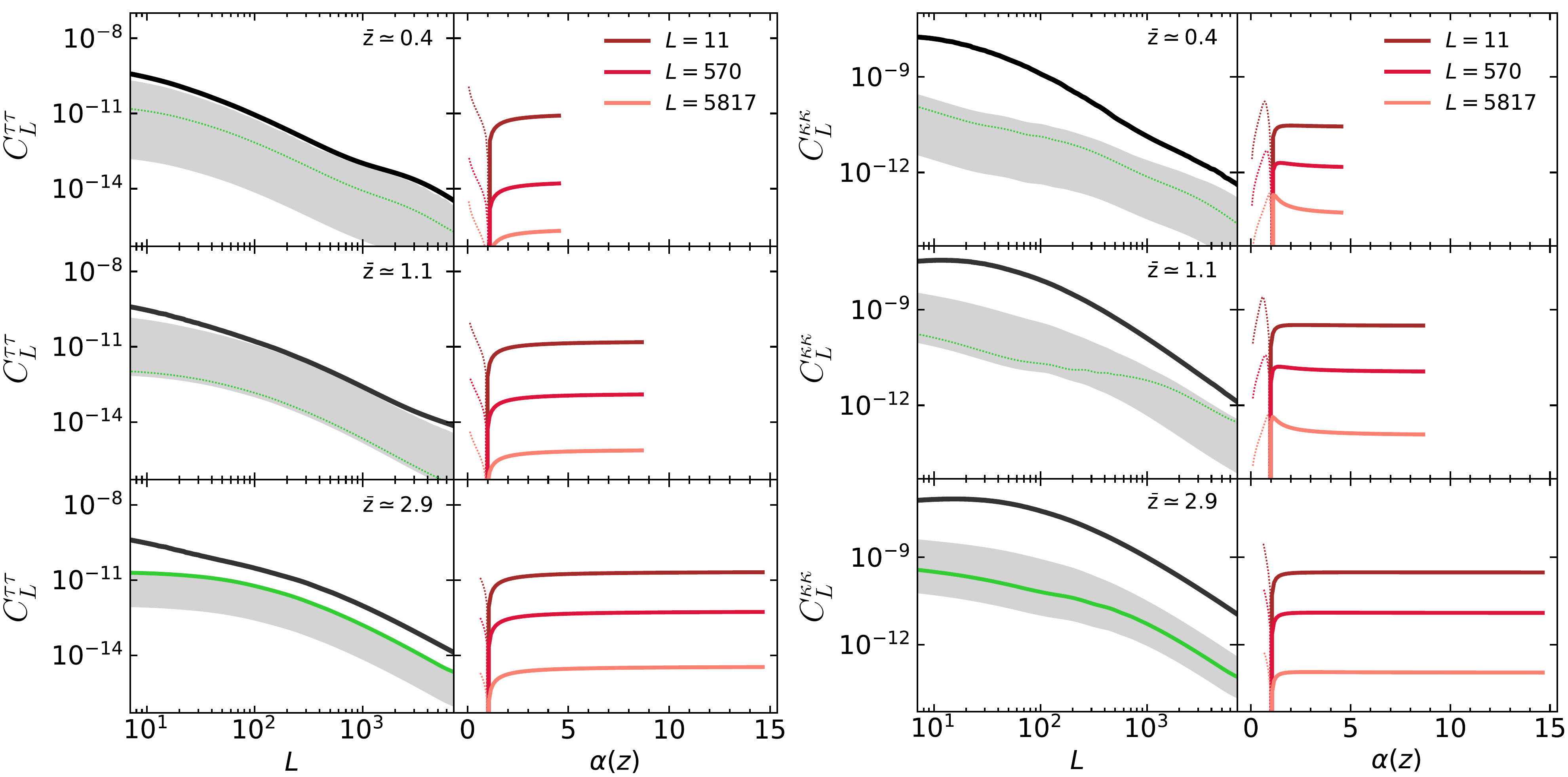}
    \vspace*{-0.4cm}
    \caption{\textit{The biases on lensing and optical-depth reconstruction.} The left set of panel shows the lensing bias to optical-depth reconstruction, where shaded regions span the region between the minimum and maximum values the bias takes for $\alpha(z)$ values varied within the same ranges described in Fig.~\ref{fig:alpha_analysis2}. The right set of panels show the effect of Thomson screening on lensing reconstruction. The left and right columns of panels in each set are otherwise similar to Fig.~\ref{fig:alpha_analysis4}. The green lines correspond to reconstruction biases $\mathcal{B}^{\tau\tau}_L$ and {$\mathcal{B}^{\kappa\kappa}_L$} assuming the LSST `gold sample' specifications introduced above.}\label{fig:alpha_analysis4}
\end{figure*}

\vspace*{-0.2cm}
\subsection{Galaxy number counts}\label{app:mag_bias_gal}
\vspace*{-0.2cm}

Fig.~\ref{fig:alpha_analysis2} demonstrates the contribution to galaxy number counts from lensing and Thomson screening. The rows of panels correspond to the three redshift bins matching Figs.~\ref{fig:clgg_biases}-\ref{fig:clpp_biases}. We vary the magnification bias between $[0.04,4.3]$, $[0.09,9.0]$ and $[0.6,15]$ for top, middle and bottom rows of panels. The left columns of panels in each set (of $3\!\times\!2$ panels)  show the galaxy power spectra in solid black. The gray-shaded regions on the same panels correspond to the region spanned by the bias contributions to galaxy power spectrum between the minimum and maximum of $\alpha(z)$ values. The right columns of panels in each set show the values of the bias contributions at fixed multipole values $\ell\in\{11,570,5817\}$ for varying $\alpha(z)$. 

The left set of panels shows the total contribution from combination of both lensing and Thomson screening (labelled `{$\kappa$} and $\tau$ combined'). For all redshifts, the bias contribution change sign at $\alpha(z)=1$ due to $\propto(1-\alpha(z))$ dependence of lensing to magnification bias. We find lensing (in particular the lensing cross-correlation terms in Eq~\eqref{eq:galaxy_total})  dominates the combined contribution. For lowest redshift and $\alpha(z)$ values below (above) unity, the total contribution is negative (positive). For higher redshifts and larger $\alpha(z)$ values, contributions to from the second row of terms in Eq~\eqref{eq:galaxy_total} become significant (in particular lensing auto-correlation) leading to bias to galaxy spectra becoming more complicated. The right set of panels shows the bias contribution from the optical depth alone. 

\vspace*{-0.2cm}
\subsection{Reconstruction noise}\label{app:mag_bias_recn}
\vspace*{-0.2cm}

The contribution of lensing and Thomson screening to the galaxy spectrum directly impacts the quadratic-estimator reconstruction noise of these observables. Fig.~\ref{fig:alpha_analysis} shows the dependence of quadratic-estimator reconstruction noise on $\alpha(z)$. The rows and $\alpha(z)$ ranges match Fig.~\ref{fig:alpha_analysis2}. 

We find the optical-depth reconstruction noise (left set of panels) depend on the magnification bias in a straight-forward way, proportional $\alpha(z)-$squared. The dependence of lensing reconstruction noise on the magnification bias however is more complicated due to competing effects from $(1-\alpha(z))$ dependent terms, and the shear-like term defined in Eq.~\eqref{eq:galaxy_realspace}. Nevertheless the former term appears more important as the reconstruction noise increases with magnification bias for values away from unity in both directions. Similar to before, the green lines correspond to the magnification bias choice used in previous sections.

\vspace*{-0.2cm}
\subsection{Reconstruction biases}\label{app:mag_bias_recsig}
\vspace*{-0.2cm}

Since their induced distortions are similar, the reconstructed lensing and optical-depth fluctuations are biased due to each other. These biases are calculated in Eqs.~\eqref{eq:bias1}~and~\eqref{eq:bias2} and displayed in Fig.~\ref{fig:alpha_analysis4}. We find both lensing and optical-depth biases to be negative (positive) for $\alpha(z)\!<\!1$ ($\alpha(z)\!>\!1$). We find lensing bias to optical depth become comparable with the underlying signal for small values of $\alpha(z)$ and lower redshifts. The bias to lensing due to Thomson screening is smaller. The $\alpha(z)$ dependence of reconstruction biases is simpler compared to galaxy spectrum.

\begin{figure*}[t!]
    \centering
    \includegraphics[width=0.96\textwidth]{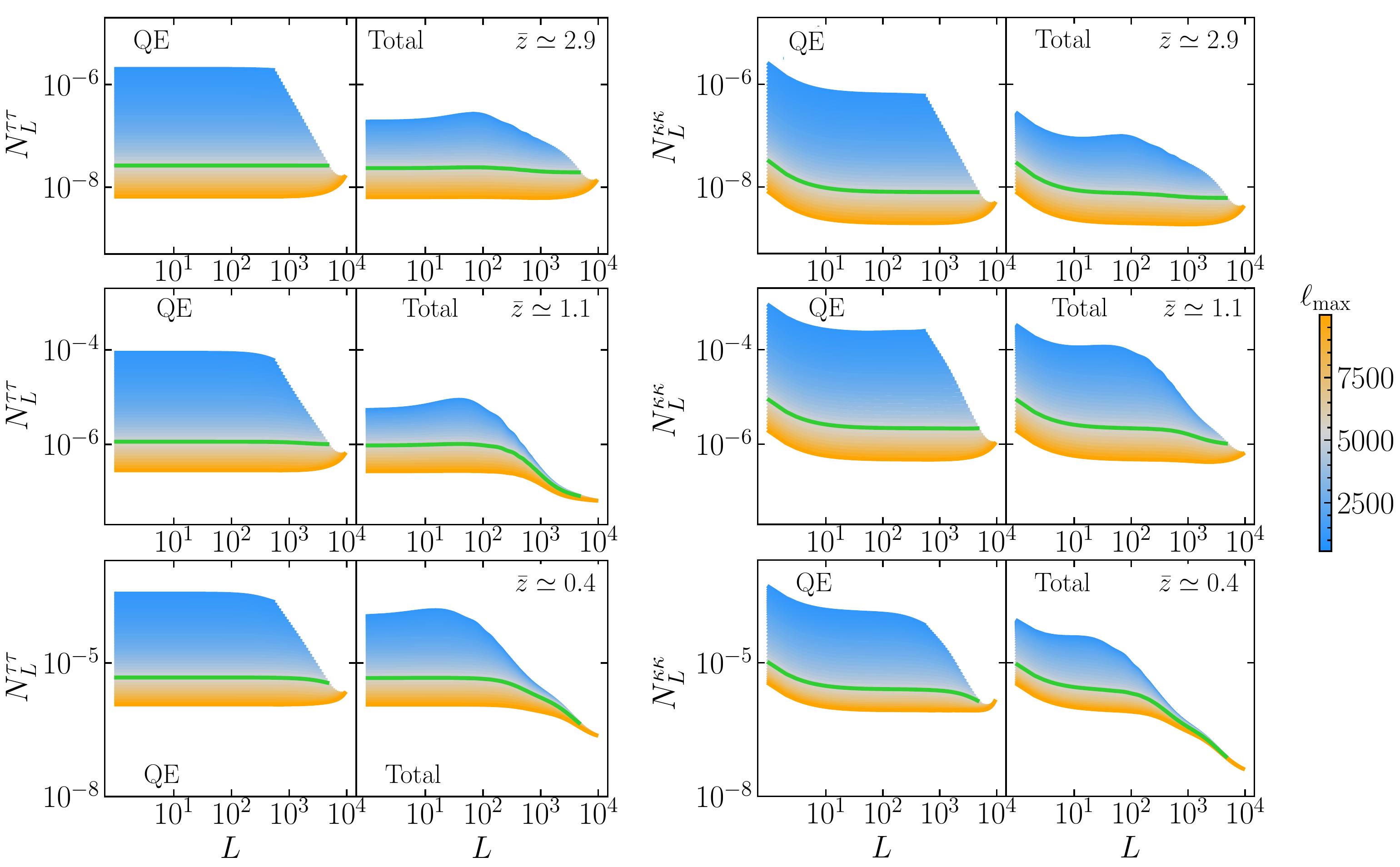}
    \vspace*{-0.4cm}
    \caption{\textit{The reconstruction noise spectra for a range of largest-multipole values assumed in the noise calculation}. The reconstruction noise are defined in Eqs.~\eqref{eq:quadnoise_tt}~and~\eqref{eq:quadnoise_pp}. The $x$-axes correspond to multipoles, $L$. The rows of panels correspond to different redshift bins defined equivalently to Figs.~\ref{fig:clgg_biases}-\ref{fig:clpp_biases}. The left (right) set of panels correspond to optical-depth (lensing) reconstruction noise. The left column of panels in each set corresponds to reconstruction noise from the quadratic estimator. The right panel corresponds to the total reconstruction noise defined in Eqs.~\eqref{eq:totalnoise_tt}~and~\eqref{eq:totalnoise_pp} by combining the linear and quadratic terms. The lines with colors ranging from blue-gray-orange correspond to $\ell_{\rm max}$ values within $[500,10000]$. Increasing $\ell_{\rm max}$ reduces the quadratic-estimator reconstruction noise. The green lines correspond to the noise values satisfying $\ell_{\rm max}=5000$ assumed in this paper.} 
    \label{fig:lmax_analysis}
    \vspace*{-0.1cm}
\end{figure*}

\begin{figure*}[t!]
    \centering
    \includegraphics[width=0.96\textwidth]{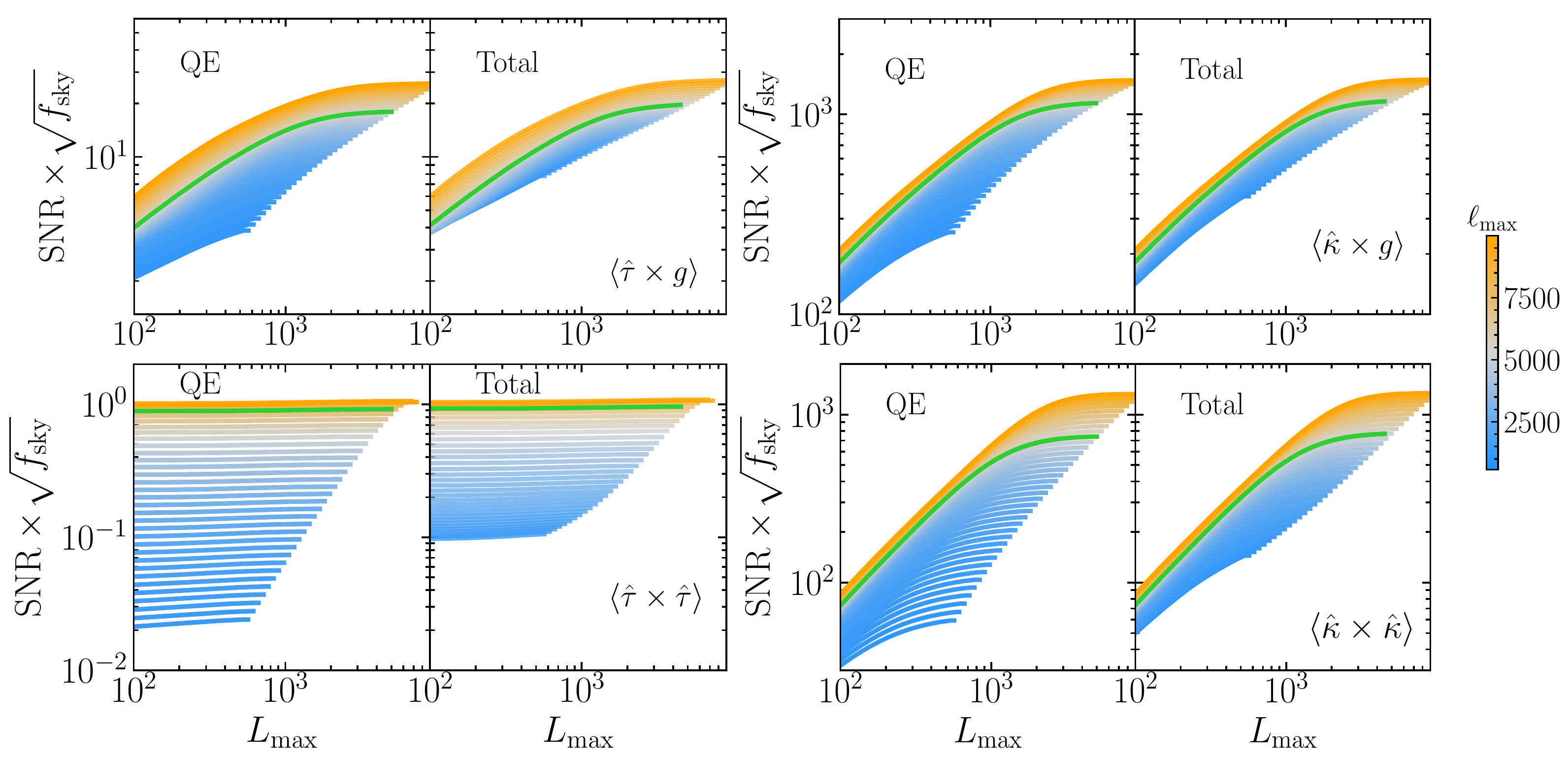}
    \vspace*{-0.3cm}
    \caption{\textit{The SNR for a range of maximum multiple values considered in the quadratic estimator.} The colors correspond to $\ell_{\rm max}$ values varied within [500,10000], similar to~Fig.~\ref{fig:lmax_analysis}. The top panels correspond to SNR values from optical-depth (left) and lensing (right) reconstruction auto-correlations. The bottom panels correspond to cross-correlations of the fields with galaxy. The left (right) subpanels in each subplot correspond to SNR calculated using the quadratic (total) estimator.}
    \label{fig:lmax_analysis_SNR}% 
    \vspace*{-0.2cm}
\end{figure*}

\vspace*{-0.2cm}
\section{The dependence of $\ell_{\rm max}$}\label{sec:appendix}
\vspace*{-0.2cm}

Both the quadratic-estimator reconstruction noise values and the SNR depends on the maximum multipole considered in the integrals defined in Eqs.~\eqref{eq:quadnoise_tt} and \eqref{eq:quadnoise_pp}. Here we study this dependence to gain further insight into the prospects of probing lensing and optical-depth from galaxy number count statistics. The ability of future surveys to use small scale fluctuations will depend on the modelling uncertainties due to non-linearity of galaxy density and baryonic effects as well as other systematics such as photo-$z$ errors and redshift-space distortions. In principle these uncertainties can be expressed in the form of reconstruction biases (in addition to those discussed in this paper), similar to the `kSZ optical-depth bias' in velocity tomography~\citep{Smith:2018bpn}, for example. We leave calculating and modelling such biases to future work. Here we demonstrate the sensitivity of our results on the $\ell_{\rm max}$ value. 

\vspace*{-0.2cm}
\subsection{Quadratic reconstruction noise}
\vspace*{-0.2cm}

In Fig.~\ref{fig:lmax_analysis} we show the dependence of optical-depth and lensing reconstruction noise to varying $\ell_{\rm max}$. The left (right) set of panels correspond to optical-depth (lensing) reconstruction. The left (right) panels in each of the sub-panels correspond to the reconstruction noise from the quadratic estimator defined in Eqs.~\eqref{eq:quadnoise_tt}~and~\eqref{eq:quadnoise_pp} (total estimator from combination of the quadratic and linear terms). Line colors correspond to different $\ell_{\rm max}$ values varied withing the range $[500,10000]$. For all cases we find around two orders of magnitude change in the noise depending on the $\ell_{\rm max}$ values. The quadratic estimators appear to dominate the (inverse) contribution to total noise at large scales ($L\lesssim100$) for $\ell_{\rm max}\gtrsim3000$, at all redshifts. The linear term improves the total noise significantly on small scales. The improvement is more enhanced at lower redshifts. Note the linear term does not depend on $\ell_{\rm max}$, and hence the differences between lines of different $\ell_{\rm max}$ values are due to changing quadratic reconstruction noise in all panels. The linear contributions to inverse variance can be seen by tracing the lowest $\ell_{\rm max}$ values of total reconstruction noise on the right panels, which are proportional to galaxy spectrum by a factor depending on the magnification bias. 

\vspace*{-0.2cm}
\subsection{Signal to noise}
\vspace*{-0.2cm}

Fig.~\ref{fig:lmax_analysis_SNR} shows the dependence of SNR on $\ell_{\rm max}$. The right (left) sub-panels correspond to SNR calculated using both linear and quadratic terms (only the quadratic term) in reconstruction. The left (right) set of panels correspond to SNR from optical-depth (lensing) reconstruction. The $y$-axes correspond to SNR normalized with $f_{\rm sky}=1$. The green lines correspond to $\ell_{\rm max}=5000$ choice made in this paper. The addition of the linear term on top the quadratic term is shown to improve the SNR significantly on all cases for $\ell_{\rm max}\lesssim3000$. The improvement from the linear contribution is marginal for large $\ell_{\rm max}$ values, similar to what is shown in Fig.~\ref{fig:lmax_analysis}. Note our results for lensing reconstruction match with the literature such as Ref.~\citep{Nistane:2022xuz}, where we find SNR of $\sim50$ from lensing reconstruction alone, with $\ell_{\rm max}=3000$ and $L_{\rm max}=1500$ setting $f_{\rm sky}=0.35$.

\bibliography{bibliography}

\end{document}